\documentclass[10pt]{article}
\usepackage{amssymb}
\usepackage{amsbsy}
\usepackage{epsfig}
\newcommand\CC {{\mathcal C}}
\newcommand\DD {{\mathcal D}}
\newcommand\FF {{\mathcal F}}
\newcommand\HH {{\mathcal H}}

\newcommand\OO {{\mathcal O}}
\newcommand\PP {{\mathcal P}}

\newcommand\TT {{\mathcal T}}
\newcommand\VV {{\mathcal V}}
\newcommand\ZZ {{\mathcal Z}}
\def\ds{\displaystyle}
\def\res{\mathop{\mathrm{res}}\limits_}

\makeatletter
\@addtoreset{equation}{section}
\makeatother
\newtheorem{theorem}{Theorem}[section]
\newtheorem{remark}{Remark}[section]
\newtheorem{proposition}{Proposition}[section]
\newtheorem{lemma}{Lemma}[section]
\newtheorem{corollary}{Corollary}[section]
\newtheorem{definition}{Definition}[section]

\def\m{\mathop}
\def\tr{\mathrm {Tr}}
\def\diag{\mathrm {diag}}
\def\BP{\boldsymbol{\Pi}}
\def\BPhi{\boldsymbol{\Phi}}
\def\BG{\boldsymbol \Gamma}
\def\BS{\boldsymbol \Sigma}
\def\BU{\boldsymbol U}

\def\ra{\rightarrow}
\def\1{{\bf 1}}
\def\br{\begin{remark}\rm\small}
\def\er{\end{remark}}
\def\wt{\widetilde}
\def\bt{\begin{theorem}}
\def\et{\end{theorem}}
\def\bd{\begin{definition}}
\def\ed{\end{definition}}
\def\bp{\begin{proposition}}
\def\ep{\end{proposition}}
\def\bl{\begin{lemma}}
\def\el{\end{lemma}}
\def\bc{\begin{corollary}}
\def\ec{\end{corollary}}
\def\be{\begin{equation}}
\def\ee{\end{equation}}
\def\bea{\begin{eqnarray}}
\def\eea{\end{eqnarray}}
\def\beaq{\begin{eqnarray}}
\def\eeaq{\end{eqnarray}}
\def \pa{\partial}
\def\Cbb{{\mathbb C}}
\def\Rbb{{\mathbb R}}
\def\Nbb{{\mathbb N}}
\def\Zbb{{\mathbb Z}}
\def\&{&{\hskip -20pt}}
\begin{document}
%
%
\font\teneuf=eufm10  \font\nineeuf=eufm9  \font\eighteuf=eufm8  
\font\seveneuf=eufm7 \font\fiveeuf=eufm5
\newfam\euffam \def\gr{\fam\euffam\teneuf}
\def\frak{\fam\euffam\teneuf}
\textfont\euffam=\teneuf \scriptfont\euffam=\seveneuf 
\scriptscriptfont\euffam=\fiveeuf
%
%
\def\grA{{\gr A}}	\def\gra{{\gr a}}
\def\grB{{\gr B}}	\def\grb{{\gr b}}
\def\grC{{\gr C}}	\def\grc{{\gr c}}
\def\grD{{\gr D}}	\def\grd{{\gr d}}
\def\grE{{\gr E}}	\def\gre{{\gr e}}
\def\grF{{\gr F}}	\def\grf{{\gr f}}
\def\grG{{\gr G}}	\def\grg{{\gr g}}
\def\grH{{\gr H}}	\def\grh{{\gr h}}
\def\grI{{\gr I}}	\def\gri{{\gr i}}
\def\grJ{{\gr J}}	\def\grj{{\gr j}}
\def\grK{{\gr K}}	\def\grk{{\gr k}}
\def\grL{{\gr L}}	\def\grl{{\gr l}}
\def\grM{{\gr M}}	\def\grm{{\gr m}}
\def\grN{{\gr N}}	\def\grn{{\gr n}}
\def\grO{{\gr O}}	\def\gro{{\gr o}}
\def\grP{{\gr P}}	\def\grp{{\gr p}}
\def\grQ{{\gr Q}}	\def\grq{{\gr q}}
\def\grR{{\gr R}}	\def\grr{{\gr r}}
\def\grS{{\gr S}}	\def\grs{{\gr s}}
\def\grT{{\gr T}}	\def\grt{{\gr t}}
\def\grU{{\gr U}}	\def\gru{{\gr u}}
\def\grV{{\gr V}}	\def\grv{{\gr v}}
\def\grW{{\gr W}}	\def\grw{{\gr w}}
\def\grX{{\gr X}}	\def\grx{{\gr x}}
\def\grY{{\gr Y}}	\def\gry{{\gr y}}
\def\grZ{{\gr Z}}	\def\grz{{\gr z}}
\def\grGl{{\gr Gl}}	\def\grgl{{\gr gl}}
\def\grSp{{\gr Sp}}	\def\grsp{{\gr sp}}
\def\grSl{{\gr Sl}}	\def\grsl{{\gr sl}}
\def\grSU{{\gr SU}}	\def\grsu{{\gr su}}
\def\nchi{\hbox{\raise 2.5pt\hbox{$\chi$}}}

\begin{flushright}
CRM-3169  (2004)\\
\hfill Saclay-T04/019
\end{flushright}
\vspace{0.2cm}
\begin{center}
\begin{Large}
\textbf{Semiclassical orthogonal polynomials, matrix models  \\
          [10pt]  and isomonodromic tau functions}\footnote{
Research supported in part by the Natural Sciences and Engineering Research Council
of Canada, the Fonds FCAR du Qu\'ebec  and EC ITH Network HPRN-CT-1999-000161.}
\end{Large}\\
\vspace{1.0cm}
\begin{large} {M.
Bertola}$^{\dagger\ddagger}$\footnote{bertola@crm.umontreal.ca}
,  { B. Eynard}$^{\dagger
\star}$\footnote{eynard@spht.saclay.cea.fr}
 and {J. Harnad}$^{\dagger \ddagger}$\footnote{harnad@crm.umontreal.ca}
\end{large}
\\
\bigskip
\begin{small}
$^{\dagger}$ {\em Centre de recherches math\'ematiques,
Universit\'e de Montr\'eal\\ C.~P.~6128, succ. centre ville, Montr\'eal,
Qu\'ebec, Canada H3C 3J7} \\
\smallskip
$^{\ddagger}$ {\em Department of Mathematics and
Statistics, Concordia University\\ 7141 Sherbrooke W., Montr\'eal, Qu\'ebec,
Canada H4B 1R6} \\ 
\smallskip
$^{\star}$ {\em Service de Physique Th\'eorique, CEA/Saclay \\ Orme des
Merisiers F-91191 Gif-sur-Yvette Cedex, FRANCE } \\
\end{small}
\bigskip
\bigskip
{\bf Abstract}
\end{center}
 The differential systems satisfied by orthogonal polynomials  with
 arbitrary semiclassical measures supported on contours in the complex
 plane are derived, as well as the compatible systems of deformation
 equations obtained from varying such measures. These are shown to
 preserve the generalized monodromy of the associated rank-$2$
 rational covariant derivative operators.  The  corresponding matrix
 models, consisting of unitarily diagonalizable matrices with spectra
 supported on these contours are analyzed, and it is shown that all
 coefficients of the associated spectral curves are given by
 logarithmic derivatives of the partition function or, more generally,
 the gap probablities. The associated isomonodromic tau functions are
 shown to coincide, within an explicitly computed factor,  with these
 partition functions. 

\tableofcontents
\section{Introduction}
The partition function for Hermitian random matrix models with
measures that are exponentials of a polynomial potential was shown in
\cite{BEH} to be equal, within a multiplicative factor independent of
the deformation parameters,  to the Jimbo-Miwa--Ueno isomonodromic 
tau function \cite{JMU} for the  rank $2$ linear differential  system satisfied 
by the corresponding set of orthogonal polynomials.  The results of 
\cite{BEH} were  in fact more general, in that  polynomials orthogonal
with respect to complex measures supported along certain contours 
in the complex plane were considered.  These may be viewed as 
corresponding to unitarily diagonalizable matrix  models in which the 
spectrum is  constrained to lie on these  contours.   

The purpose of the  present  work is to extend these considerations to the 
more general setting of complex  measures whose logarithmic derivatives 
are  arbitrary {\em rational} functions, the associated {\it semiclassical} 
orthogonal polynomials and generalized matrix models. By also including 
contours with endpoints, the latter viewed as further deformation
parameters, the gap probability densities are included as special cases of
partition functions.

 To place the results in context, we first briefly recall the main
 points of \cite{BEH}, restricting to the more standard case of
 Hermitian matrices and real measures. Consider orthogonal polynomials 
 $\pi_n(x) \in L^2(\Rbb,{\rm e}^{-\hbar^{-1}V(x)}dx)$  supported on 
 the real line, with the measure defined by  exponentiating a real 
 polynomial potential 
\be
V(x) = \sum_{J=1}^{d} \frac {t_J}  J x^J\ .
\ee
 (Here we assume $V(x)$ is of even degree and with positive leading
coefficient, although  these restrictions are unnecessary in the more 
general setting of \cite{BEH}.) The small parameter $\hbar$ is  usually 
taken as  $\OO( N^{-1})$ when considering the limit $N\to \infty$. 

Any two consecutive polynomials satisfy a first order system of ODE's  
\be
\hbar \frac {d}{dx} \pmatrix {\pi_{n-1}(x)\cr \pi_n(x)} =
 \DD_n(x) \pmatrix {\pi_{n-1}(x)\cr \pi_n(x)}  ,
\label{uno}
\ee
where $\DD_n(x)$ is a $2 \times 2$ matrix with  polynomial coefficients of 
degree at most $d-1=\deg(V'(x))$. The infinitesimal deformations corresponding 
to changes in the coefficients $\{t_J\}$  result in a sequence of
Frobenius compatible, overdetermined systems of PDE's 
\beaq
\hbar \frac{\pa}{\pa t_J}  \pmatrix {\pi_{n-1}(x)\cr \pi_n(x)} = T_{n,J}(x)
\pmatrix {\pi_{n-1}(x)\cr \pi_n(x)}   \quad J=1, \dots , d ,\label{due}
\eeaq
where the matrices $T_{n,J}(x)$ are  polynomials in
$x$ of degree $J$ which satisfy the compatibility conditions 
\beaq
\left[\hbar \frac{\pa}{\pa t_J}-T_{n,J}(x),\hbar  \frac{d}{ dx} -\DD_n(x)
  \right]=0\ .
\eeaq
 It follows that the generalized monodromy data of the sequence of
 rational covariant derivative operators $ \hbar \frac{d}{d x} -\DD_n(x)$  
 are invariant under these deformations, and independent of  the integer $n$.  
 
 This is a particular case of the general problem of rational isomonodromic
deformation systems  \cite{JMU}.  An important r\^ole is
played in this theory by the  {\em isomonodromic  tau function}
$\tau_n^{IM}$ associated with any  solution of an isomonodromic
deformation system. This function on the space of deformation
parameters is obtained by integrating a closed differential whose
coefficients are given by residues involving the fundamental solutions
of the system. The main results of \cite{BEH} were the following. First,  
the coefficients of the associated {\it spectral curve}, given by the 
characteristic  equation of the matrix $\DD_n(x)$, can be obtained by 
applying certain first order differential operators   with respect to the deformation 
parameters  (Virasoro generators) to $\ln (\ZZ_n(V))$, where 
the partition function $\ZZ_n(V)$ of the associated $n \times n$  matrix model 
is 
\be
\label{Zndef}
\ZZ_n(V):= \int_{\HH_n} dM\,\exp\left(-\frac 1 \hbar\tr V(M)\right)  \ .
\ee
Second,  this partition function  is equal to the isomonodromic tau function up to 
a multiplicative factor that does not depend on the deformation parameters
\be
\label{Ztaupolynom}
\tau_n^{IM} = \ZZ_n(V)\,\FF_n .
\ee

The present work generalizes these results to the case of  measures whose
logarithmic derivatives are  arbitrary rational functions, including
those supported on curve segments in which the endpoints may play the
r\^ole of further deformation parameters. The latter are of importance
in the calculation of gap probabilities in matrix models \cite{TW}
since these may, in this way, using measures supported on such segments 
 be put on the same footing as partition functions \cite{BS}. A Frobenius 
compatible system of first order differential and deformation equations 
satisfied by the corresponding orthogonal polynomials is derived 
(Propositions \ref{PropVfolding}, \ref{Propdiffrecfold}) and the coefficients 
of the associated spectral curve are again shown to be obtained by 
applying suitable Virasoro generators to $\ln (\ZZ_N(V))$  (Theorems
\ref{spectral}--\ref{eval}). A  formula that generalizes  (\ref{Ztaupolynom}) 
is also derived (Theorem \ref{main} ): 
\be
\tau_n^{IM} = \ZZ_n(V)\FF_n(V)  \ ,
\ee
where the factor $\FF_n(V)$ is an explicitly computed  function of the
deformation parameters determining  $V$, which can in fact be eliminated by
a making a suitable scalar gauge transformation.

 The results  of Theorems \ref{spectral}, \ref{eval}  give a precise meaning, for 
 finite $n$, to formul\ae\      
 that are usually derived  in  the asymptotic limit $n\to\infty$ ($\hbar n \sim \OO(1)$)
 through saddle point computations, relating the free energy to the
 asymptotic spectral curve. It is well known \cite{DGZ} that  the free
 energy in the large $n$ limit  is given by solving a minimization
 problem (in the Hermitian matrix model) 
\be
\mathcal F_0:= - \lim_{n\to\infty} \hbar^2 \ln \ZZ_n
 =\min_{\rho(x)\geq 0} \left[\int V(x)\rho(x)dx -  \int\int
 \rho(x)\rho(x') \ln|x-x'| \right]\ ,
\ee
giving the equilibrium density $\rho_{\rm eq}$ for the eigenvalue distribution.
If, for instance,  the potential is a real polynomial  bounded from below,   it is
 known \cite{DKMVZ} that the support of the equilibrium density is
 a union of finite segments $I\subset \Rbb$. The  density $\rho_{\rm eq}$ is obtained from the 
 variational equation 
 \be 
2 \PP \int{\rho_{\rm eq}(x) dx \over x-x' }= V'(x)
 \label{vareqn}
 \ee
(within $I$), and is related  to the {\em resolvent} by
\be
\omega(z):=\hbar \lim_{n\to \infty} \left< \tr {1\over M -  z }\right> =\int_Idx \frac {\rho_{\rm eq}(x)}{x-z}\ ,
\ \ z\in
 \Cbb\setminus I\ .
\ee
In terms of this, the spectral density may be recovered  as the
 jump-discontinuity  of $\omega(z)$ across $I$, and all its moments are given by 
\be
\pa_{t_J}\mathcal F_0 = \frac 1 J \lim_{n\to\infty}\hbar\left\langle \tr
 M^J\right\rangle = \int_I dx \frac {x^J}{J}\rho_{\rm eq}(x) =
 - \res{z=\infty} \frac {z^J}J \omega(z) dz \ .  \label{strnzt}
\ee
The function $y= -\omega(x)$  satisfies an algebraic relation  given by 
\be
y^2 = yV'(x) + R(x)\ ,
\label{infinspecurve}
\ee 
where $R(x)$ is a polynomial of
degree less than $V'(x)$ that is  uniquely determined by  the consistency 
of  (\ref{vareqn}) and
(\ref{infinspecurve}). 
 
The point  to be stressed here is that  this asymptotic spectral curve
should be compared with the  spectral curve of Theorem \ref{spectral},
given by the   characteristic equation (\ref{spectralcurven}), which also  
contains all the relevant information about the  finite $n$ case.
 In the $n\rightarrow \infty$ limit,  logarithmic derivatives of the  partition 
 function  are expressed in  (\ref{strnzt}) as  residues of the meromorphic
differentials $z ^k ydz$ on the curve. The {\it same} formulae are shown
in Theorem \ref{eval}  to hold as {\it exact} relations for the finite $n$ case
if we replace the ``asymptotic''  spectral curve by the spectral curve given
 by the characteristic equation of the  matrix $\DD_n(x)$.
 
The paper is organized as follows.
In section \ref{Orthopol}  the problem is defined in terms of
polynomials orthogonal with respect to an arbitrary semiclassical
measure supported on complex contours, and the corresponding
generalized matrix model partition functions.  In section \ref{Derivs}
the recursion relations, differential systems and deformation
equations which these satisfy are expressed  in terms of the
semi-infinite ``wave vector'' formed from the orthogonal polynomials.
In section  \ref{Folded}  the notion of ``folding'' is introduced and
used  (Propositions \ref{PropVfolding}--\ref{Propdiffrecfold}) to
express the preceding equations as an infinite sequence of compatible
overdetermined $2 \times 2$ systems of linear differential equations
and recursion relations satisfied by pairs of consecutive orthogonal
polynomials. In section \ref{Spectral} the results of folding  are
used to express the spectral curve in terms of logarithmic derivatives
of the partition  function and it is shown  that the $n\to
\infty$ relation between the free energy and the spectral curve is
also valid as an exact result for finite $n$. In section
\ref{Isomonodromic} the definition of the isomonodromic tau function
\cite{JMU} is recalled and it is computed by relating it to the
spectral invariants of the rational matrix generalizing $\DD_n(x)$ in
(\ref{uno}). These invariants are shown to give the logarithmic
derivatives of the tau functions in terms of residues of  meromorphic
differentials  on the spectral curves through formaulae that are nearly identical to
those for the partition function, This leads to the main result,
Theorem \ref{main}, which gives the explicit relation between
$\ZZ_n$ and $\tau_n^{IM}$. 

\section{Generalized orthogonal polynomials and partition functions}
\label{Orthopol}

\subsection{Orthogonality measures and  integration contours}
 Given a measure on the real line, the associated orthogonal
 polynomials are those that  diagonalize the quadratic form associated
 to the corresponding (complex) {\em moment functional}; i.e.,  the
 linear form obtained  by integration with respect to the measure. 
 \be
\begin{array}{rcl}
\mathcal L:\Cbb[x]&\to& \Cbb\ .\cr
p(x)&\mapsto& \mathcal L(p(x)) = \int_{\Rbb} p(x) d\mu(x)\ .
\end{array}
\ee
A natural generalization consists of including moment functionals that
are expressed by integration along more general contours in the
complex $x$ plan,  with respect to a complex measure defined by
locally analytic weight functions that may have isolated essential singular
points and complex power-like branch points.  

We thus consider linear forms on polynomials given by integrals of the form
\beaq
\mathcal L(p(x)) &\&= \int_\varkappa p(x) \mu(x)dx \cr
\mu(x)&\& = {\rm e}^{-\frac 1 \hbar V(x)} \cr
 V(x) &\&:= \sum_{r=0}^{K} T_r(x)\ ,\label{mu}
\eeaq
where 
\beaq
T_0(x)&\&:=t_{0,0}+ \sum_{J=1}^{d_0}\frac {t_{0,J}}J x^J\cr
T_r(x)&\&:=  \sum_{J=1}^{d_r} \frac
  {t_{r,J}}{J(x-c_r)^J} -t_{r,0}\ln(x-c_r)\cr
&\&-\hbar \pa_x\ln\mu(x)= V'(x)  = \sum_{J=1}^{d_0 -1} t_{0,J}
x^{J-1} - \sum_{r=1}^{ K} \sum_{J=1}^{d_r+1} \frac
{t_{r,J-1}}{(x-c_r)^J} \ ,
\eeaq
and the symbol $\int_\varkappa $ denotes integration over linear combinations of contours on which the integrals
are convergent, as explained below. 
This class of linear functionals is sometimes referred to as {\em semiclassical} moment functionals \cite{B, MR1,MR2}.
We consider  the corresponding monic generalized orthogonal polynomials $p_n(x)$,  which satisfy 
\be
\int_\varkappa p_n(x)p_m(x) \mu(x)dx =h_n\delta_{nm} \ .
\label{orthopol}
\ee
 If all the contours are contained in the real axis and the weight is
 real and positive, we reduce to the usual notion of semiclassical
 orthogonal polynomials. 
The small parameter $\hbar$ introduced in (\ref{mu}) is not of
 essential importance here; it is only retained in the formul\ae\
 below to recall that,  when 
taking the  large $n$ limit,  it plays the r\^ole of small parameter
 for which $\hbar n$  remains finite as  $n\to\infty$.
\par
To describe the contours of integration,
we first define sectors $S_r^{(j)}$, $r=0,\dots K$, $k=1,\dots d_r$.
around the points $c_r$ for which $d_r>1$ ($c_0:=\infty$) in such a way that 
\be
\Re\left(V(x)\right) \mathop{\longrightarrow}_{
\shortstack{{\scriptsize $x\to c_r,$}\\
{\scriptsize  $x\in S
^{(j)}_r$}}}
+\infty\ .
\ee
The number of sectors for each pole in $V$ is equal to the degree of that pole; that
is, $d_0 $ for the pole at infinity and $d_r$ for the  pole at $c_r$.
Explicitly 
\beaq
S^{(0)}_k &\&:= \left\{ x\in \Cbb;\ \ \frac {2k\pi
 -\arg(t_{0,d_0})-\frac \pi 2}{d_0}< \arg(x) < \frac {2k\pi
 -\arg(t_{0,d_0}) +\frac \pi 
2}{d_0} \right\} ,\label{sectinf}\\
&\&\ k=0\dots d_0-1\ ; \cr
&\&S^{(r)}_k := \left\{ x\in \Cbb;\ \ \frac {2k\pi  +\arg(t_{r,d_r})-\frac \pi
2}{d_r}< \arg(x-c_r) < \frac {2k\pi +\arg(t_{r,d_r}) +\frac \pi
2}{d_r} \right\},\\
&\& k=0,\dots,d_r-1,\ \ r=1,\dots,K\ . \nonumber
\eeaq
These sectors are defined in such a way that approaching any
of the essential singularities of $\mu(x)$ (i.e. a $c_r$ such that $d_r>0$) 
within them, the function $\mu(x)$ tends to zero faster than any power.\\

\subsubsection{Definition of the boundary-free contours} 
The definition of the contours follows \cite{Mi} (see fig. $1$).
\begin{enumerate}
\item For any $c_r$ for which there is {\em no essential singularity}
  in the measure (i.e. $d_r=0$), there are  two subcases: 
\begin{enumerate}
\item For  the $c_r$'s that are  branch points or poles in $\mu$ (i.e.,  $t_{r,0}\notin \Nbb$), we  
take a loop starting at infinity in some fixed sector
$ S_{k}^{(0)}$  encircling the singularity and going back to infinity in the same
sector. (Note that if $c_r$ is just a pole; i.e., $t_{r,0}\in -\Nbb^+ $, the contour could 
equivalently be taken as a circle around $c_r$.)
\item For the $c_r$'s  that are regular points ($t_{r,0} \in \Nbb$ ),  we
take a line joining $c_r$ to infinity, approaching $\infty$ in a sector $S^{(0)}_{k}$
  as before. 
\end{enumerate}
\item For any $c_r$ for which there is an essential singularity in
  $\mu$  (i.e. , $d_r >0$) we define $d_r$ contours starting from $c_r$
in the sector $S_k^{(r)}$ and returning to it  in the next    sector $S_{k+1}^{(r)}$. 
Also, if $t_{r,0}\notin  \Zbb$, we join the singularity $c_r$ to $\infty$ by a path 
approaching $\infty$ within one fixed sector $ S^{(0)}_{k}$.
\item For $c_0:=\infty$, we take $d_0-1$ contours starting at $c_0$ in the
sector $S^{(0)}_k$ and returning at $c_0$ in the next sector
$S^{(0)}_{k+1}$.
\end{enumerate}
Note that, with these definitions, the integrals involved are
convergent and  we can perform integration by parts.
Moreover, any contour in the complex plane for which the integral of $\mu(x)p(x)dx$
is convergent for all polynomials $p(x)$ is equivalent to a linear combination
of the contours defined above, no two of which are, in this sense, equivalent.

\subsubsection{Definition of the hard-edge contours}

We also include some additional contpours in the complex plane 
$\{m)j\}_{j=1, \dots, L}$,   starting at  some points $a_{j}$, $j=1\dots L$ 
and going to $\infty$ within one of the sectors $S_{k}^{(0)}$. These 
could be viewed as corresponding  to additional points in 1(b) for
which both $d_r=0$ and $t_{r,0}=0$, but we prefer to deal with them
separately since integration by parts on these contours does give a contribution.

In total there are $S:= d_0 + \sum_{r=1}^{K} (d_r+1) $ boundary-free contours
$\sigma_\ell,\ ,\ell=1,\dots,S$ and $L$ hard-edge contours $m_h,\ h=1,\dots,L$. 
The moment functional is an arbitrary linear combination of integrals taken along these
contours
\be
\int_\varkappa :=\sum_{j=1}^{L}\varkappa_{j}\int_{m_j}+ \sum_{j=1}^S\varkappa_{L+j}\int_{\sigma_j}  \ \ .
\ee
Note that, by taking appropriate linear combinations of the
 contours, we could alternatively have had  contours consisting of finite
segments joining the points $a_j$.\\
\centerline{
\parbox{8cm}{
\epsfxsize=8cm
\epsfysize=8cm
\epsffile{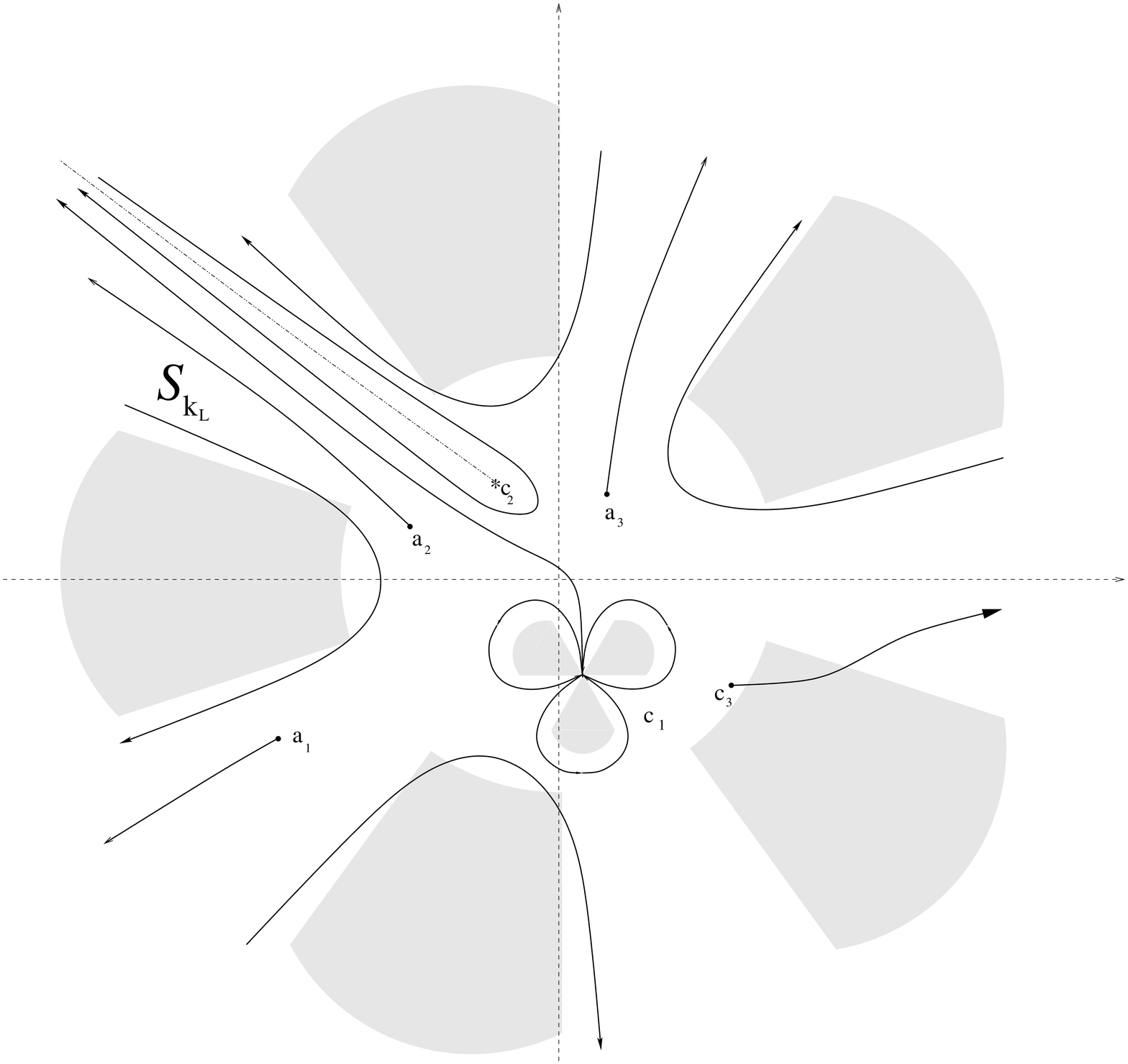}}}
\centerline{ \parbox{12cm}{Figure 1: The types  of contours considered in the $x$ Riemann sphere
$\mathbb P^1$. Here we have $c_1$ with $d_1=3$ and $c_2$ with $d_2=0,
    t_{2,0}\notin \Zbb$  (logarithmic singularity in the potential), $c_3$ with $d_3=0$, $t_{3,0}\in \Nbb$
and the degree of the potential at infinity $c_0=\infty$ is $d_0=5$.
 The essential singularity in $\mu$  at $c_1$ is of the form
 $\exp{(x-c_1)^{-3}}$ and there is also a cut extending from $c_1$ to
 $\infty$ if $t_{1,0}\notin  \Zbb$. The point $c_2$ is a branch point
 of  $\mu(x)$ since $t_{2,0}\notin \Zbb$, and the cut extends to
 infinity ``inside'' the contour (as shown here). If it were a pole
 ($t_{2,0}\in -\Nbb^+$), the contour would be replaced by a circle
 around  it. The point $c_3$ is a regular point with $t_{3,0}\in
 \Nbb^\times$, and the contour extending from it to infinity is no
 different from the ones starting at the regular points $a_1, a_2,
 a_3$.  The latter are the ``hard-edge'' segments joining the points
 $a_1,a_2$ and $a_3$ to  $\infty$ within one of the sectors
 $S^{(0)}_{k}$. 
}}

\subsection{Recursion relations, derivatives  and deformations equations}
\label{Derivs}
\subsubsection{Existence of orthogonal polynomials and relation to random matrices}
Recall  \cite{Ch} that orthogonal polynomials  satisfying
(\ref{orthopol}) exist provided all the Hankel determinants formed
from the moments are nonzero: 
\be
\Delta_n(\varkappa):= \det\left[\int_\varkappa x^{i+j} \mu(x){\rm
    d}x\right]_{0\leq i,j\leq n-1}\neq 0\ ,\ \ \forall n\in \Nbb.\label{hankel}
\ee
Since the $\Delta_n(\varkappa)$'s are homogeneous polynomials in the
coefficients $\varkappa_j$, the zero locus excluded by (\ref{hankel})
is of zero measure  (in the space of $\varkappa_j$'s), and hence
``generically'' the conditions (\ref{hankel}) are fulfilled. 
The development to follow will in fact only involve  orthogonal
polynomials up to some arbitrarily large fixed degree, say $N$, and
hence the conditions $\Delta_n(\varkappa)=0\ ,n\leq N-1$ determine a
Zariski closed set in $\{\varkappa_j\}$, (and a closed set of measure
zero in the space of 
coefficients of $V$). 

The orthogonal polynomials considered here are related to models of
unitarily diagonalizable random matrices $M\in gl(n,\Cbb)$ with
spectra  supported  on the contours defined above.  More specifically
we have the partition function
\beaq
 \ZZ_n&\&:=C_n\int_{spec(M)\in \varkappa} dM {\rm
   e}^{-\frac 1 \hbar \tr
  V(M)}  \cr 
&\&=\int_{\varkappa}{\rm
d}x_1\cdots\int_{\varkappa}dx_n 
\, \Delta(\underline x)^2 {\rm e}^{-\frac 1 \hbar\sum_{j=1}^n V(x_j)} \cr
&\&=
n!\Delta_n(\varkappa,V)  =n! \prod_{j=0}^{n-1} h_j\  ,
\label{partit}
\eeaq
where
\be
C_N :={1 \over \left(\int_{U(n)}dU\right)}
\ee
is the inverse of the $U(n)$ group volume, and
\be
\Delta(\underline x):= \prod_{i<j}
(x_i-x_j)\label{vandermonde}
\ee
is the usual Vandermonde determinant. The notation $spec(M)\in \varkappa$ in the first integral
just means 
\be
M=U\,D\,U^\dagger\ , \qquad D:= {\rm diag} (x_1, \dots x_n) \ , \qquad U\in U(n) \ ,
\ee
where the eigenvalues $\{x_1, \dots x_n\}$ of $M$ are constrained to lie on the contours entering in $\int_\varkappa$.

  In particular, as in the standard case,  the orthogonal polynomials
  may be shown to be equal to the expectation values of the
  characteristic polynomials in such models 
  \beaq
  p_n(x) &\&= \left < \det(x {\bf I} - M)\right>  \cr
  &\&={1 \over Z_n} \int_{\varkappa}{\rm
d}x_1\cdots\int_{\varkappa \BG}dx_n 
\,  \prod_{i=1}^n(x-x_i)\Delta(\underline x)^2 {\rm e}^{-\frac 1 \hbar\sum_{j=1}^n V(x_j)}  \ ,
  \eeaq
  and all correlation functions between the eigenvalues may be expressed
  as determinants in terms of the standard Christoffel-Darboux kernel formed
  from them
  \be
  K_n(x,y) := \sum_{j=0}^{n-1}{1\over h_j}p_j(x)p_j(y)e^{-{1\over 2\hbar}(V(x) +V(y))}\ .
  \ee
More precisely, this  is valid when there are no ``hard-edge''
  contours present. Inclusion of the latter  however allows one to
  interpret these determinants as certain conditional correlators,
  known as ``Janossy distribution"  correlators \cite{BS},  giving the
  probability densities for a certain number of eigenvalues to lie at
  given locations within the complementary part of the support, while
  the remaining ones lie within it. The partition function $Z_n$ in
  this case can be reinterpreted as the corresponding gap probability
  \cite{BS, TW}. 
  
\subsubsection{Wave vector equations}

We now define  the normalized  orthogonal polynomials
\be
\pi_n(x):= \frac 1 {\sqrt{h_n}} p_n(x)
\ee
  and what will be referred to as  the  `` orthonormal quasi-polynomials'' 
\be
\psi_n(x):=\pi_n(x){\rm e}^{-\frac 1 {2\hbar} V(x)} ,
\ee
satisfying
\be
\int_\varkappa\psi_n(x) \psi_m(x)  d x = \delta_{mn}.
\ee
From the former, we form the  semi-infinite  ``wave vectors''
\be
{\BP}(x):= [\pi_0(x),\pi_1(x),\dots,\pi_n(x),\dots]^t\ .
\ee

As in the theory of ordinary orthogonal polynomials, we have
\be
x\BP(x)=  Q \BP(x) \label{recs}\ ,
\ee
where $Q$ is a  symmetric  tridiagonal semi-infinite matrix with components
\be
Q_{ij} = \gamma_j \delta_{i, j-1}  + \beta_i \delta_{ij} + \gamma_{i} \delta_{i, j+1}, \quad i, j \in \Nbb \ ,
\ee
defining a three term recursion relation of the form 
\be
x\pi_j(x) = \gamma_{j+1}\pi_{j+1}(x) + \beta_j \pi_j(x) + \gamma_j \pi_{j-1}(x)\ .
\ee

Now introduce semi-infinite matrices $P, A_i, C_r, T_{r,J}$  such that 
\bea
\hbar \pa_x\BP(x) &\&= P\BP(x)  \label{deformP}\\
\hbar\pa_{a_i}\BP(x) &\&= A_i\BP(x)\ ,\qquad i=1, \dots , L\label{deformA}\\
\hbar\pa_{c_r}\BP(x)&\& = C_r\BP(x)\ ,\qquad r=1, \dots , K \label{deformC}\\
 \hbar\pa_{t_{r,J}}\BP(x) &\&= T_{r,J}\BP(x)\ ,\qquad r=0, \dots , K,\
J=0,  \dots , d_r   .\label{deformT}
\eea
Their matrix elements are determined simply by integration
\be
X_{nm} = \int_\varkappa\left(\hbar \pa \pi_n(x) \right)\pi_m(x) \mu(x)dx\ ,
\ee
where $\pa$ denotes any of the derivatives $\pa_x,\pa_{a_i},
\pa_{c_r}, \pa_{t_{r,J}}$ above for which $X$  becomes the
corresponding matrices $P$, $A_i$, $C_r$ or $T_{r,J}$ on the RHS of 
(\ref{deformP}) - (\ref{deformT}).

\br
Such wave vectors and associated deformation equations have been
studied in many previous works relating orthogonal polynomials, matrix
models and integrable systems (see, e.g. \cite{AvM2},
\cite{vM}). However, considerations of the deformation theory have
mainly been within the formal setting, with the potential  $V(x)$
replaced by some initial value, $V_0(x)$, plus a perturbation
consisting of an infinite power series with arbitrary coefficients,
without regard to domains of convergence. Results  obtained in this
formal setting cannot be directly applied to the study of
isomonodromic deformations, where the local analytic structure in the
neighborhood of  a number of isolated singular points is of primary
interest. 

\er
For any such semi-infinite square matrix  $X$,  let $X_0$, $X_+$,
$X_-$ denote the diagonal, upper and lower triangular parts,
respectively, and let 
\be
X_{-0} := \frac 1 2 X_0  + X_-\ .
\ee
\bp
\label{DeformeqsPi}
The matrices $P$, $A_i $, $C_r$ and $T_{r,J}$ are all lower
semi-triangular (with $P$ strictly lower triangular) , and are given
by 
\beaq
P &\& = V'(Q)_{-0} - \sum_{i=1}^{L} A_i =  V'(Q)_{-} - \sum_{i=1}^{L}(A_i)_- 
 \label{relation}  \\
A_i &\&=\hbar\varkappa_{i} (\BP(a_i)\BP^t(a_i))_{-0}  \label{Adef} \\
C_r &\&= -\sum_{J=0}^{d_r} t_{r,J} (Q-c_r)^{-J-1}_{-0} \ ,  \quad r=1, \dots  , K \label{Cdef}\\
 T_{0,0} &\&= \frac 12 \1  \label{T00def}\\
T_{0,J}&\& = \frac 1 {J} Q^J_{-0}\ ,\quad J=1,\dots, d_0 \label{T0jdef}\\
T_{r,J} &\&= \frac 1 {J} (Q-c_r)^{-J}_{-0}, \quad r=1, \dots  ,K, \quad J= 1, \dots ,  d_r
\label{Trjdef} \\
T_{r,0} &\&=  -\ln(Q-c_r)_{-0}, \quad r=1, \dots  , K \ .
\label{Tr0def} 
\eeaq
where   $(Q-c_r)^{-J}$ and $\ln(Q-c_r)$ are defined by the formul\ae\    
\bea
 (Q-c_r)^{-J}_{nm}&\&:= \int_\varkappa \frac {\pi_n(z)\pi_m(z)}{(z-c_r)^J}\mu(z)dz\\
\ln(Q-c_r)_{nm}&\&:= \int_\varkappa \ln(z-c_r)\pi_n(x)\pi_m(z) \mu(z)dz\ .
\eea
The diagonal matrix elements for each of the above is given by the formula
\be
\label{diagonalX}
X_{jj} = -{\hbar\over 2} \pa(\ln h_j) \ ,
\ee
where $\pa = \pa_x$, $\pa_{a_i}$,
$\pa_{c_r}$ and $\pa_{t_{r,J}}$, respectively, for $X  = P$, $A_i$,
$C_r$ and $T_{r,J}$.   In particular, they vanish for $P$, which is
strictly lower triangular, and hence 
\be
\label{diagVprime}
V'(Q)_{jj} = \hbar \sum_{i=1}^{L} \kappa_i \psi_j(a_i)^2 \ .
\ee
\ep
{\bf Proof.} 
We make use of the orthogonality relations
\be
\int_\varkappa \BP(x)\BP^t(x){\mu}(x)dx = \1  \label{orthogPi}  \ .
\ee
Eqs. (\ref{Cdef}) - (\ref{Tr0def}) are obtained as follows. 
 Consider a  deformation $\pa$  with respect to any of the above
 $c_r$'s or $t_{r,J}$'s   and denote by $X$ the corresponding matrix;
 then 
\beaq
\hbar\pa \pi_n(x) &\&=\hbar \pa \frac {p_n(x)}{\sqrt{h_n}}  
= -\frac 1 2 \left(\hbar\pa  \ln(h_n)\right) \pi_n(x) +
\frac 1 {\sqrt{h_n}}\hbar \pa p_n(x)   \label{defdiag}\\
&\&= -\frac 1 2 \left(\hbar\pa
  \ln(h_n)\right) \pi_n(x) +\hbox{ lower degree polynomials}\ ,
\eeaq
since the polynomials $\pi_n$ are monic. It follows that the  deformation matrix $X$ is lower semi-triangular.
On the other hand, differentiating eq.~(\ref{orthogPi}) gives
\be
0 = \int_\varkappa \left(\hbar\pa\BP\,\BP^t + \BP\,\hbar\pa \BP^t\right)\mu(x)dx +
\int_\varkappa \BP\,\BP^t\hbar \pa\mu(x)dx = X+X^t + \int_\varkappa \BP\,\BP^t \hbar\pa\mu(x)dx
\ee
Applying the operators for each case to $\mu$ as defined in (\ref{mu})
and using eq.~(\ref{recs}) then gives the result.

Now consider the deformations of the endpoints $a_i$ of the ``hard-edge'' contours.
Differentiating (\ref{orthogPi}) gives
\beaq\
0 &\&=\hbar \pa_{a_i}\int_\varkappa\BP\BP^t\mu(x) dx =-
\varkappa_{i} \hbar \BP(a_i)\BP^t(a_i)\mu(a_i)
+ \int_\varkappa\left[ ( \hbar\pa_{a_i}\BP)\BP^t + \BP\hbar\pa_{a_i}\BP^t\right]\mu(x)dx \cr
&\&=-\hbar  \varkappa_{i} \BP(a_i)\BP^t(a_i) \mu(a_i)+ A_i+A_i^t ,
\eea
where 
\be
A_i:= \hbar \int_\varkappa \pa_{a_i} \BP\, \BP^t \mu (x)dx  \ ,
\ee
It follows that 
\be
A_i =\hbar\varkappa_{i} (\BP\BP^t)_{-0}\big|_{x=a_i} \ ,
\ee
proving eq. (\ref{Adef}), and also that
\be
(A_i)_{nn} =-{\hbar \over 2} \pa_{a_i}\ln(h_n) 
 ={\hbar \over 2}\varkappa_{i} \psi_n^2(a_i)  \ .
 \label{Ainnlnpsi}
\ee

To determine the matrix $P$, note that it is {\em strictly lower triangular} and  
\beaq
\hbar\BP\BP^t\bigg|_{\pa \varkappa} &\&=-\hbar\sum_{i=1}^{L}
\left(A_i+A_i^t\right) =\hbar\int_\varkappa \left(\BP'\,\BP^t +
\BP\,\BP'^t 
+\BP\,\BP^t\pa_x\ln\mu(x) \right)\mu(x) dx  \cr
&\&= \int_\varkappa \left ( \hbar\BP'\,\BP^t + \hbar \BP\,\BP'^t
-V'(x) \BP\,\BP^t \right)\mu(x) dx
= P+ P^t  - V'(Q).\label{PP}
\eeaq
This implies that  
\be
P =  V'(Q)_{-0} - \sum_{i=1}^{L} A_i =  V'(Q)_{-} - \sum_{i=1}^{L}(A_i)_-  \, .
\ee
This last equality  follows from  (\ref{diagVprime}), which, in turn, follows
from  integration by parts in the definition of $V'(Q)_{nn}$. It may be
seen as a consequence of the  invariance of the partition function
under an infinitesimal change in the integration variables $x_j
\rightarrow x_j +\epsilon$ in (\ref{partit}); i.e., translational
invariance. 

   From  (\ref{diagonalX}) and (\ref{partit})  follows a relation
   between the diagonal elements of the  deformation matrices and the
   logarithmic derivatives of the partition function that will be
   very important in what follows. Define the truncated trace of a
   semi-infinite matrix $X$ to be 
\be
\tr_n X:=\sum_{j=0}^{n-1} X_{jj} \ .
\ee  
\bc
For $\pa= \pa_{a_j} $, $\pa_{c_r}, $ and $\pa_{t_{r,J}}$, 
\be
\hbar\pa\ln\ZZ_n = -2 \tr_n X \ ,
\ee
with $X  = A_j$, $C_r$ and $T_{r,J}$, respectively. 
For the cases $\pa_{c_r} $ and $\pa_{t_{r,J}}$,
\be
\hbar\pa\ln\ZZ_n = \tr_n \pa V(Q)\ ,
\ee
while for the $\pa_{a_i}$'s we have
\be
\sum_{i=1}^{L} \kappa_i \pa_{a_i}\ln \ZZ_n = -\hbar V'(Q)_{nn}
\ee
\ec
\noindent{\bf Proof.} 
 The first of these relations follows from (\ref{diagonalX}) and
 (\ref{partit})  directly, the second from the explicit expressions
 for the deformation matrices (\ref{Cdef})--(\ref{Tr0def})  and of the
 potential $V(x)$, and the third is a restatement of the  
(\ref{diagVprime}) (translational invariance).
\bc
The compatibilty conditions
\be
\left[\mathcal G, \HH\right] =0\ ,
\ee
are satisfied, 
where $\mathcal G, \ \HH$ are any of the following operators
\be
 \hbar\pa_{a_i}-A_i,\ \  \hbar\pa_{t_{r,J}} - T_{r,J},\ \ \hbar \pa_{c_r} - C_r ,   \ \ \hbar\pa_x - P,\ \ x-Q 
\ee
and  $r=0,\dots K$, $J=0,\dots d_r$.
\ec
{\bf Proof.} 
This follows immediately from the fact that the orthogonal
polynomials entering in eqs. (\ref{deformP})-(\ref{deformT}) are linearly independent. 
\br
\label{Laxremark}
Note that  
\be
[ \hbar\pa_x - P, x-Q] =0
\ee
 is just  the string equation, while the other compatibility conditions involving $x-Q$ 
 imply the Lax  equations:
 \be
 \hbar \pa_{a_i} Q = [A_i, Q], \qquad 
  \hbar \pa_{t_{r,J}} Q = [T_{r,J}, Q], \qquad
   \hbar \pa_{c_r} Q = [C_r, Q] \ ,
   \ee
   showing that  the spectrum of the matrix $Q$ is invariant under these deformations.
   \er

 \subsubsection{Wave vector of the second kind}
 
We now consider solutions of the second kind, 
\be
\phi_n(x) := {\rm e}^{\frac 1 \hbar V(x)}\int_\varkappa \frac {{\rm
    e}^{-\frac 1 \hbar V(z)}\pi_(z)}{x-z}dz . \label{second} \ ,
\ee
which may be combined to form the components of a wave vector of 
the second kind
\be 
\BPhi(x) := [\phi_0(x),\phi_1(x),\dots,\phi_n(x),\dots]^t\ .
\ee

Denote by  
\be
\nabla_QV'(x) := \frac {V'(x)-V'(Q)}{x-Q}
\ee 
the semi-infinite square matrix with elements
\be
\left(\frac {V'(x)-V'(Q)}{x-Q}\right)_{nm}  = \int_\varkappa dz {\rm
  e}^{-\frac 1 \hbar V(z)}\pi_n(z)\pi_m(z)  \frac {V'(x)-V'(z)}{x-z} \ ,
\ee
 and define $\BU(x)$ to be the semi-infinite column vector (with only its zeroth 
 component nonvanishing) given by
\be
(\BU(x))_n := \sqrt{h_0} {\rm  e}^{\frac 1 \hbar V(x)}\delta_{n,0} .  
\ee

The following lemma gives the effect of multiplication  of $\BP(x)$ by
$x$ and of application of $\hbar \pa_x$ to it. It may be deduced
immediately from eqns. (\ref{recs}) and (\ref{deformP}), applied
inside the integral, together with integration by parts. 
\bl
\label{QPPhi}
\bea
x\BPhi(x) &\&= Q\BPhi(x) +\BU(x)
 \label{recsPhi}\\
\hbar \pa_x\BPhi(x) &\&= P\BPhi(x) +
  \nabla_Q V'(x) \BU(x) +\hbar \sum_{i=1}^{L}\varkappa_{i}\frac {{\rm
       e}^{\frac 1 \hbar \left(V(x)-V(a_i)\right)}}{x-a_i}\BP(a_i)  \ ,\label{deformPhi}
\eea
\el

The next proposition, which is similarly verified, gives the effects
of the above deformations on the wave vector of the second kind. 

\bp
\label{DeformeqsPhi}
\bea
\hbar \pa_{a_i}\BPhi(x) &\&= A_i\BPhi(x)-\hbar \varkappa_{i} \frac {{\rm
  e}^{\frac 1 \hbar(V(x)-V(a_i))}}{x-a_i} \BP(a_i), \quad i=1, \dots, L,   \label{PhideformA}\\
\hbar  \pa_{c_r} \BPhi(x) &\&= C_r \BPhi(x) + \sum_{J=0}^{d_r} t_{r,J} \frac
    {(Q-c_r)^{-j-1}-(x-c_r)^{-j-1}}{Q-x} \BU(x), \quad r=1, \dots , K,  \label{PhideformC}\\
   \hbar \pa_{t_{0,J}} \BPhi(x) &\&= T_{0,J} \BPhi(x) + \frac {Q^J-x^J}{Q-x} \BU(x) ,  \
   quad J= 1, \dots , d_0, 
    \label{PhideformT0}\\
\hbar  \pa_{t_{r,J}}\BPhi(x) &\&= T_{r,J} \BPhi(x) + \frac
    {(Q-c_r)^{-J}-(x-c_r)^{-J}}{Q-x} \BU(x) \ ,\quad J= 1, \dots , d_r, \quad r=1, \dots, K,
         \label{PhideformTr} \\
\hbar  \pa_{t_{r,0}}\BPhi(x) &\&= T_{r,0} \BPhi(x)  + \frac
    {\ln(Q-c_r)-\ln(x-c_r)}{Q-x} \BU(x),  \quad r = 1, \dots , K.
    \label{PhideformTr0}
\eea
\ep

    The content of eqs.~(\ref{deformC})--(\ref{deformT}) and
    (\ref{PhideformC})--(\ref{PhideformTr0}) may be summarized
    uniformly as follows. 
Let $v(x)$ be any function that is analytic at each point of the contours except,
possibly,  the points $c_r$, and for which the following integrals are
convergent:
\bea
v(Q)_{nm} &\&:= \int_\varkappa v(z) \pi_n(z)\pi_m(z){\rm e}^{-\frac 1 \hbar V(z)}dz= \int_\varkappa
v(z) \psi_n(z)\psi_m(z)dz  \cr
(\nabla_Q v(x))_{nm}&\&:= \left(\frac {v(x)-v(Q)}{x-Q}\right)_{nm} := \int_\varkappa
 \frac {v(x)-v(z)}{x-z}\psi_n(z)\psi_m(z)dz .
 \label{nablaQvdef}
 \eea
Define the deformation matrix  under the infinitesimal variation of the
potential $V(x)\mapsto V(x) + v(x)$ to be 
\be
X_v:= v(Q)_{-0}\ .
\ee
Then the two infinite systems
\bea
\delta_v \BP(x) &\&:= X_v\BP(x)\label{vvarPi}\\
\delta_v \BPhi(x) &\&:= X_v\BPhi(x) +  \nabla_Q v(x) \BU(x)
\label{vvarPhi}
\eea
describe the infinitesimal deformation of the orthogonal polynomials
and the second-kind solutions under such infinitesimal variations of the
potential.

Equivalently, define the $2\times \infty$ matrix 
\be
\BG(x):= \left[\BP(x),\BPhi(x)\right] .
\ee
 In terms of $\BG(x)$, all the recursion, differential and deformation
 equations (\ref{recs}), (\ref{deformP})--(\ref{deformT}) and
 (\ref{recsPhi})--(\ref{PhideformTr0}) may be expressed simultaneously
 as 
\bea
x\BG &\&= Q \BG + \left( 0, \BU\right)  
\label{recsGamma}\\
\hbar \pa_x \BG &\&= P\BG + \left(0, \nabla_Q V' \BU  + \hbar
\sum_{i=1}^K {e^{{1\over \hbar} (V(x) - V(a_i))} \over x
  -a_i}\BP(a_i)\right) 
\label{dPGamma}\\
\delta_v\BG &\&= X_v \BG + (0, \nabla_Q v \BU) 
\label{dvGamma}\\
\hbar \pa_{a_i} \BG &\&= A_i \BG - \left(0, \hbar\varkappa_i
  {e^{{1\over \hbar} (V(x) - V(a_i))} \over x -a_i}\BP(a_i)\right),
\label{daiGamma}
\eea
where $v$ signifies any of the infinitesilmal deformations of the
potential $\hbar \pa_{c_i}, \hbar\pa_{t_{r,J}}V$ ($i=1, \dots L, \  r=
0, \dots K, \ J=1, \dots , d_r$).

\section{Folding}
\label{Folded}
\subsection{$n$-windows and Christoffel-Darboux formula}

Let $i_n$ be the $\infty \times 2$  matrix that represents the injection of the 2-dimensional
subspace spanned by the $(n-1, n)$ basis elements into the (semi-)infinite space corresponding
to the components of $\Psi$ or $\BPhi$. Its matrix elements are thus:
\be 
(i_n)_{jk} = \delta_{k,1} \delta_{j, n-1} + \delta_{k,2}\delta_{j, n}, \quad j = 0, 1, 2, \dots,  \quad k=1,2.
\ee
Let $i_n^T$ denote its transpose, which is the corresponding projection operator.
The  $n$-th $2\times 2$  block (or ``{\bf window}'') of $\BG$ is then given by:
\be
\BG_n(x) := i_n^T \BG = \left[\begin{array}{cc}
\pi_{n\!-\!1}(x) & \phi_{n\!-\!1}(x)\\
\pi_{n}(x) & \phi_{n}(x)
\end{array}\right] \ .
\ee

By ``folding'' the infinite recursion and differential-deformation equations
(\ref{recs}), (\ref{deformP})--(\ref{deformT}),
(\ref{recsPhi})--(\ref{PhideformTr0}),
(\ref{recsGamma})--(\ref{daiGamma}),  we mean the corresponding
sequence of recursion relations, ODEs and PDEs satisfied by the
$\BG_n(x)$'s. To derive these, a  form of the Christoffel--Darboux
identity for orthogonal polynomials will repeatedly be used. 
Let $\BP_n^0$  denote the semi-infinite square matrix whose only
nonvanishing entries are $1$'s on the diagonal in positions $0$ to $n$
(i.e. the projection onto the first n+1 components) 
\be
(\BP_n^0)_{ij} := \left\{ \begin{array}{cc}  \delta_{ij} & \quad {\rm if} \quad 0 \le i,j \le n \\
                                            0 &\quad {\rm otherwise} \ . quad
                                \end{array} \right. 
\ee
Let
\be 
\sigma:= \left( \begin{array} {cc}0 & -1 \\ 1 &  \ \ 0 \end{array}\right)
\ee
be the standard $2 \times 2$ symplectic matrix, and let 
\be
\BS_n := i_n \sigma i_n^T\label{symplec}
\ee
denote its projection onto the $2\times 2 $ subspace in position $(n-1, n)$.

\bp\label{modifiedCD}
The following {\bf extended Christoffel-Darboux} formul\ae\     are satisfied:
\bea
&\&(x-x') \BG^T(x) \BP_{n-1}^0 \BG(x')  \\ 
&\&  =\gamma_n \BG_n^T(x) \sigma \BG_n(x')    +\left ( \begin{array}{cc}   0  & - e^{{1\over \hbar}V(x') } \\
      e^{{1\over \hbar}V(x)}   & 
 e^{{1\over \hbar} (V(x) + V(x'))} \int_{\varkappa} e^{-{1\over 2}
   V(z)} \left( {1\over x-z} - {1\over x'-z}\right) dz 
    \end{array} 
      \right)   \\
&\&  = \gamma_n \BG^T(x) \BS_n \BG(x')    + \left ( \begin{array}{cc}   0  & - e^{{1\over \hbar}V(x') } \\
      e^{{1\over \hbar}V(x)}   
          &  e^{{1\over \hbar} (V(x) + V(x'))} \int_{\varkappa}
 e^{-{1\over 2} V(z)} \left( {1\over x-z} - {1\over x'-z}\right)  dz 
    \end{array} 
      \right)   \ .
      \label{extendedCD}
         \eea
      Equivalently, in components,
      \bea
      (x-x')\sum_{j=0}^{n-1}\pi_j(x)\pi_j(x')  &\&=
\gamma_{n}\left[\pi_n(x)\pi_{n-1}(x') - \pi_{n-1}(x)\pi_{n}(x')\right]
\label{CD11} \\
  (x-x')\sum_{j=0}^{n-1}\pi_j(x)\phi_j(x')  &\&=
\gamma_{n}\left[\pi_n(x)\phi_{n-1}(x') - \pi_{n-1}(x)\phi_{n}(x')\right] - {\rm
  e}^{\frac 1 \hbar V(x')}\\
    (x-x')\sum_{j=0}^{n-1}\phi_j(x)\phi_j(x')  &\&=
\gamma_{n}\left[\phi_n(x)\phi_{n-1}(x') - \phi_{n-1}(x)\phi_{n}(x')\right] \cr
&\& +e^{{1\over \hbar} (V(x) + V(x'))} \int_{\varkappa} e^{-{1\over 2} V(z)} \left( {1\over x-z} - {1\over x'-z}\right) dz ,
\eea
and evaluating at $x=x'$ gives
\be
\det \BG_n(x) = \pi_{n-1}(x)\phi_{n}(x) - \phi_{n-1}(x)\pi_{n}(x)
=-\frac 1{\gamma_n} {\rm e}^{\frac 1 \hbar V(x)}\
  \label{CDdet} .
\ee
\ep
{\bf Proof.}  Eq. (\ref{CD11}) is the standard Christoffel-Darboux
relation for orthogonal polynomials. The extended system may be
derived as follows. Multiplying  the expression $\BG(x) \BP_{n-1}^0
\BG(x')$ by $(x-x')$, and applying the relation (\ref{recsGamma}) with
respect to both $x$ and $x'$ gives 
\bea
(x-x') \BG^T(x) \BP_{n-1}^0 \BG(x') &\&= \BG^T(x) \left(Q\BP_{n-1}^0 - \pi_{n-1}^0 Q\right) \BG(x')\\
&\&  +\left ( \begin{array}{cc}   0  & - e^{{1\over \hbar}V(x') } \\
      e^{{1\over \hbar}V(x)}   &
          \quad e^{{1\over \hbar} (V(x) + V(x'))} \int_{\varkappa}
	  e^{-{1\over 2} V(z)} \left( {1\over x-z} - {1\over
	    x'-z}\right) dz  
    \end{array} 
      \right) \ .
      \eea
    The result (\ref{extendedCD}) is obtained by substituting the following  
    identity, which holds for any tridiagonal symmetric matrix of the form $Q$
     \be
     Q\BP_{n-1}^0 - \BP_{n-1}^0 Q = \gamma_n i_n \sigma i_n^T  \ .  \label{commutQident}
     \ee

\subsection{Folded version of the deformation equations for changes in the potential}

Under infinitesimal changes of the parameters in the potential $V$ and
 the end-points of the ``hard-edge'' contours,  the wave vectors
 $\BP(x)$ and $\BPhi(x)$ and the combined system $\BG(x) $ undergo changes
 determined by equations (\ref{deformA})-(\ref{deformT}), 
 (\ref{PhideformA})-(\ref{PhideformTr0})) and  
 (\ref{daiGamma})-(\ref{dvGamma}). Besides the deformations induced by
 infinitesimal changes of the endpoints  $\{a_j\}$, all
 these deformations have the same general form, depending only on the
 function 
 $v(x) =\delta V(x)$ that gives the infinitesimal deformation of the
 potential. We deal with them all on the same footing in the following
 proposition, which expresses the explicit form they take on the
 window $\BG_n(x)$. 

\bp
\label{PropVfolding}
 The deformation equations (\ref{vvarPi}), (\ref{vvarPhi})
 (\ref{dvGamma}) are equivalent to the infinite sequence of $2\times
 2$ equations 
\be
\delta_v \BG_n(x) = \VV_n (x) \BG_n (x)
\label{folded} ,
\ee
where  the folded  matrix of the deformation is defined by  
\bea
\label{foldedVdef}
  \VV_n(x) = \left[\begin{array}{cc}
v(x)\!-\!\frac 1 2v(Q)_{n\!-\!1,n\!-\!1}\!\!\!\!\!\!\!\! & 0 \\
0 & \frac 12 v(Q)_{nn}
\end{array}\right] + \gamma_n\left[
\begin{array}{cc}
\nabla_Q v(x)_{n\!-\!1,n\!-\!1} & \nabla_Q v(x)_{n\!-\!1,n}\\
 \nabla_Q v(x)_{n,n\!-\!1} & \nabla_Q v(x)_{nn}\end{array}
\right]
\sigma \ . \label{foldeddef}
\eea
For the deformations in
(\ref{deformP})--(\ref{deformT}) and
(\ref{PhideformA})--(\ref{PhideformTr0}), this  gives the following
equations corresponding to changes in the potential. 
\bea
\hbar \pa_{c_r}\BG_n(x) &\&=\CC_{r;n}(x)\BG_n(x)  \label{Cdefn}\\
\hbar \pa_{t_{r,J}}\BG_n(x)&\&= \TT_{r,J;n}(x)\BG_n(x) , \label{Trsdefn}
\eea
where the sequence of $2\times 2$ matrices $\CC_{r;n}$ and
$\TT_{r,J;n}(x)$ are rational in $x$, with poles at the points
$\{c_r \}$, obtained by making the following substitutions in
eq.~((\ref{foldedVdef}). 
\beaq
C_r:  \ v(x)&\ra & -\sum_{J=0}^{d_r} t_{r,J} (x-c_r)^{-J-1}\cr
T_{r,J}: \  v(x)&\ra & \frac  1 J (x-c_r)^{-J}\cr
T_{0,J}: \  v(x)&\ra& \frac 1 J x^J \cr
T_{r,0}: \  v(x)&\ra &  -\ln(x-c_r)\ .
\eeaq

\ep
{\bf Proof}.
Using the definition (\ref{nablaQvdef}) of $\nabla_Q v(x)$ and the
extended Christoffel-Darboux relation (\ref{extendedCD}), we have 
\bea
\gamma_n \nabla_Qv(x) \BS_n\BG  &\&= \gamma_n \int_{\varkappa} dy e^{-{1\over \hbar} V(y)} 
\left(v(y) - v(x)\right) \BP(y) {\BP^T(y) \BS_n^T \BG(x) \over y-x}\\
 &\&=  \int_{\varkappa} dy e^{-{1\over \hbar} V(y)}  \left(v(y) - v(x)\right)  
 ) \BP(y) \BP^T(y)\BP_{n-1}^0 \BG(x) \\
 &&{\hskip -10 pt}+ \left(0,\  -e^{{1\over \hbar}V(x)} \int_{\varkappa} dy e^{-{1\over \hbar} V(y)} \left( {v(y) - v(x)\over y-x}  
 \right) \BP(y) \right) \\
 &\&= v(Q) \BP_{n-1}^0 \BG(x) - v(x) \BP_{n-1}^0 \BG(x) -  \left(0, \nabla_Q v(x) \BU(x)\right).
\eea
Applying the projector $i_n^T$ and noting that
\be
i_n^T v(Q) \BP_{n-1}^0 \BG = i_n^T v(Q)_{-0}\BG 
+ {1\over 2} \pmatrix{v(Q)_{n-1, n-1} & 0 \cr 0 &  - v(Q)_{n, n} } \BG_n,
\ee
we obtain
\bea
\hbar \delta_v \BG_n (x)  &\&= i_n^T \hbar \delta_v \BG(x) = i_n^T\left( X_v \BG(x)  + (0,  \nabla_Qv \BU(x)\right) \\
 &\&= i_n^T v(Q)_{-0} \BG (x) + (0, i_n^T \nabla_Q v(x)  \BU(x)) \\
 &\&= i_n^T v(Q)\BP_{n-1}^0 \BG (x)  + 
  \pmatrix{-  {1\over 2}v(Q)_{n-1, n-1} & 0 \cr 0 &   {1\over 2}  v(Q)_{n, n} } \BG_n+ (0, i_n^T \nabla_Q v(x)  \BU(x)) \\
 &\&= \gamma_n i_n^T \nabla_Q v(x) i_n^T \sigma \BG_n(x) + 
 \pmatrix{ v(x) - {1\over 2} v(Q)_{n-1, n-1}  & 0 \cr
 0 &  {1\over 2} v(Q)_{n, n}}\BG_n(x),
\eea
proving the relation (\ref{foldedVdef}). 

\br
Note that formula (\ref{foldeddef}) for the deformation of the measure in
Proposition \ref{PropVfolding}, as well as  those below,
(\ref{fdd}), (\ref{formulafordd}), which are obtained through folding
of the $\hbar \pa_x$ operator, could also be derived for arbitrary
locally analytic  potentials $V(x)$, provided all the integrals
involved are convergent \cite{Is}.  However applicability of the
subsequent isomonodromic analysis  would be lost if the derivatives
were not rational, since the resulting deformation equations would
then have essential singularities. 
\er

\subsection{Folding of the endpoint deformations}

The case (\ref{deformA}) and (\ref{PhideformA}) involving deformations
of the locations of the ``hard edge'' end--points must be considered
separately. 
\bp
\label{PropArdefn} 
The following gives a closed system for the $n$-th window of
eqs.~(\ref{deformA}) and (\ref{PhideformA})
\be
\hbar \pa_{a_i}\BG_n(x) = {\mathcal A}_{i,n}(x)\BG_n(x)
\label{Ardefn}
\ee
where
\bea
  {\mathcal A}_{i,n} &\&:= \frac {\hbar\varkappa_{i}\gamma_n }{a_i-x} \left[\begin {array}{cc}
\psi_{n-1}(a_i)\psi_n(a_i) &  - \psi_{n-1}^2(a_i)\\
\psi_{n}^2(a_i) & -\psi_{n-1}(a_i)\psi_n(a_i) 
\end{array}\right] + \frac {\hbar\varkappa_{i}} 2  \left[\begin {array}{cc}
-\psi_{n-1}^2(a_i)& 0\\ 0& \psi_n^2(a_i)
\end{array}\right]   \cr
&\&=  \frac {\hbar \varkappa_{i} \gamma_n}{a_i-x} \left[\begin {array}{cc}
 \psi_{n-1}^2(a_i) & \psi_{n-1}(a_i)\psi_n(a_i)  \\
 \psi_{n-1}(a_i)\psi_n(a_i)&\psi_{n}^2(a_i) 
\end{array}\right] \sigma
+ \frac {\hbar\varkappa_{i}} 2  \left[\begin {array}{cc}
-\psi_{n-1}^2(a_i)& 0\\ 0& \psi_n^2(a_i)
\end{array}\right] 
\eea
\ep
{\bf Proof. }
This is very similar to the proof of Prop. \ref{PropVfolding}. Using
 the definition  (\ref{Adef}) of the matrices $A_i$  and the extended
 Christoffel--Darboux relation  (\ref{extendedCD}) we have
\bea
 \hbar\pa_{a_i} \BG_n(x) &\&=\hbar i_n^T \pa_{a_i}\BG(x) = i_n^T\left[
\left(\hbar\varkappa_i\Psi(a_i)\Psi^T(a_i)\right)_{-0} \BG(x)   -
\left(0,\hbar\varkappa_i \frac {{\rm
  e}^{\frac 1\hbar (V(x)-V(a_i))}}{x-a_i} \BP(a_i)\right)
\right]  \cr
&\&=i_n^T\left[
\left(\hbar\varkappa_i\Psi(a_i)\Psi^T(a_i)\right) \BP_{n-1}^0  \BG(x)
- \left(0,\hbar\varkappa_i \frac {{\rm
  e}^{\frac 1\hbar (V(x)-V(a_i))}}{x-a_i} \BP(a_i)\right)
\right] \cr
&\& +\frac {\hbar\varkappa_i}2\left[ \begin{array}{cc}
-\psi_{n-1}^2(a_i) & 0\cr
0& \psi_{n}^2(a_i)
\end{array}\right]\BG_n(x) \cr
&\&= i_n^T\frac { \gamma_n \hbar\varkappa_i}{a_i-x}
\Psi(a_i)\Psi^T(a_i)\BS_n\BG(x)  +\frac {\hbar\varkappa_i}2\left[ \begin{array}{cc}
-\psi_{n-1}^2(a_i) & 0\cr
0& \psi_{n}^2(a_i)
\end{array}\right]\BG_n(x)\ .
\eea 
Recalling the definition  (\ref{symplec}) of $\BS_n$ and computing the
matrix product yields the result in the statement. Q.E.D.

\subsection{Folded version of the recursion relations and $\hbar\pa_x$  relations}

We now consider the recursion relations (\ref{recs}), (\ref{recsPhi})
and (\ref{recsGamma}) and the action of the $\hbar\pa_x$ operator in
(\ref{deformP}),  (\ref{deformPhi})  and (\ref{dPGamma}) which, in
their folded form are given by the following. 
\bp
\label{Propdiffrecfold}
The folded forms of the relations (\ref{recsGamma}) and (\ref{dPGamma})  are 
\bea
\BG_{n+1}(x)&\&=R_n(x)\BG_n(x)\ , \quad n\ge1,\label{Recn} \\
\pa_x \BG_n(x) &\&= \DD_n(x)\BG_n(x)  \label{Pdefn} 
\eea
where 
\be
R_n :=\left[
\begin{array}{cc}
0 & 1\\
-\frac {\gamma_n}{\gamma_{n+1}} & \frac {x-\beta_n}{\gamma_{n+1}}
\end{array}
\right] \ .
\ee
and 
\be
\DD_n(x) =
\DD_n^{(0)}(x)  + \sum_{i=1}^{L} \frac {\hbar \varkappa_{i}\gamma_n }{x-a_i} \left[\begin {array}{cc}
\psi_{n-1}(a_i)\psi_n(a_i) &  - \psi_{n-1}^2(a_i)\\
\psi_{n}^2(a_i) & -\psi_{n-1}(a_i)\psi_n(a_i) 
\end{array}\right]\label{fdd} 
\ee
with
\beaq
\DD_n^{(0)}(x) \!\!\!&\&=\!\!\!    \left[\matrix{ V'(x) & 0 \cr 0 & 0}\right] +
\left[\matrix{ \left(\nabla_Q V'(x)\right)_{n-1,n-1} & \left(
\nabla_Q V'(x)\right)_{n-1,n} 
\cr \left( \nabla_Q V'(x)\right)_{n,n-1} &   \left(
\nabla_Q V'(x)\right)_{nn}  }\right] \, \left[\matrix{ 0 & -\gamma_n \cr
\gamma_n & 0 }\right]     \cr
\!\!\!&\&=\!\!\!    \left[\matrix{ V'(x)  & 0 \cr 0 & 0} \right]+
\gamma_n 
 \left[\matrix{\left(
\nabla_Q V'(x)\right)_{n-1,n}  &-  \left( \nabla_Q V'(x)\right)_{n-1,n-1}\cr
\left(\nabla_Q V'(x)\right)_{nn} &  
-\left( \nabla_Q V'(x)\right)_{n,n-1}}\right]\label{formulafordd}
\eeaq
\ep

\br Note that formula (\ref{fdd}) implies that
\be
{\rm Tr}(\DD_n(x)) = V'(x)\ .
\label{trd1}
\ee
\er

\noindent {\bf Proof:}
 The folded form (\ref{Recn}) of the recursion relations follows directly from eqs.~(\ref{recs}) and (\ref{recsPhi}))\bea
x \pi_n(x) &\&= \gamma_{n+1}\pi_{n+1}(x)+\beta_n\pi_n(x) +
\gamma_n \pi_{n-1}(x)\\
x\phi_n(x) &\&= \gamma_{n+1}\phi_{n+1}(x)+\beta_n\phi_n(x) +
\gamma_n\phi_{n-1}(x) + \delta_{n0} \sqrt{h_0} {\rm e}^{\frac 1 \hbar V(x)} \ .
\eea

To prove  (\ref{Pdefn}),
 note that the folding relations
 (\ref{folded}) 
may be expressed
\be
i_n^T \delta_v \BG_n = \VV_n \BG_n
\ee
for any infinitesimal variation $v= \delta V$ in the potential. Choosing
\be
\delta := -\sum_{r=1}^K \pa_{c_r}  + \sum_{J=1}^{d_0} j t_{0,J+1}
 \pa_{t_{0, J}} + t_{0,1}\pa_{t_{0,0}} ,
\ee
we have 
\be
V'(x) \equiv \delta V .
\ee
Using (\ref{relation}) and (\ref{deformPhi}), we have
\bea
\hbar \pa_x \BG_n 
&& {\hskip -20 pt}= i_n^T \left[ P\BG +
\left( 0 \ , \ \nabla_Q  V'(x) \BU   -  \hbar  \sum_{i=1}^{L}\varkappa_{i}\frac {{\rm
       e}^{\frac 1 \hbar \left(V(x)-V(a_i)\right)}}{x-a_i}\BP(a_i)\right)\right]\\
 && {\hskip -20 pt}=  i_n^T \left[ (V'(Q)_{-0} - \sum_{i=1}^{L} A_i)\BG +
  \left( 0 \ , \ \nabla_Q  (\delta V)(x) \BU -  \ \hbar  \sum_{i=1}^{L}\varkappa_{i}\frac {{\rm
       e}^{\frac 1 \hbar \left(V(x)-V(a_i)\right)}}{x-a_i}\BP(a_i)\right)\right]\\
        && {\hskip -20 pt}=  i_n^T \left[ ((\delta V)(Q)_{-0} - \sum_{i=1}^{L} A_i)\BG +
\left( 0 \ , \ \nabla_Q  (\delta V)(x) \BU -  \ \hbar  \sum_{i=1}^{{L}}\varkappa_{i}\frac {{\rm
       e}^{\frac 1 \hbar \left(V(x)-V(a_i)\right)}}{x-a_i}\BP(a_i)\right)\right]\\
&& {\hskip -20 pt}=  i_n^T \left[ \delta - \sum_{i=1}^{L} \pa_{a_i}\right] \BG ,
 \eea
 where we have used the deformation equations (\ref{deformA})--(\ref{deformT}), (\ref{PhideformA})--(\ref{PhideformTr0}).
 Applying the folded relations (\ref{folded}), (\ref{foldedVdef}) and (\ref{Ardefn}), this gives
 \be 
 \hbar \pa_x \BG_n  
 = \left[ \VV_n - \sum_{i=1}^K \hat{{\mathcal A}}_{i,n} - \sum_{i=1}^K \bar{{\mathcal A}}_{i,n} \right] \BG_n,
 \label{VnAn}
\ee
where
\bea
\hat{{\mathcal A}}_{i,n}  &\&:= 
\frac {\hbar\varkappa_{i}\gamma_n }{a_i-x} \left[\begin {array}{cc}
\psi_{n-1}(a_i)\psi_n(a_i) &  - \psi_{n-1}^2(a_i)\\
\psi_{n}^2(a_i) & -\psi_{n-1}(a_i)\psi_n(a_i) 
\end{array}\right]   \\
\bar{{\mathcal A}}_{i,n}  &\&:= 
+ \frac {\hbar\varkappa_{i}} 2  \left[\begin {array}{cc}
-\psi_{n-1}^2(a_i)& 0\\ 0& \psi_n^2(a_i)
\end{array}\right]  ,
\eea
and
\bea 
  \VV_n(x) &\&= \left[\begin{array}{cc}
V'(x)\!-\!\frac 1 2V'(Q)_{n\!-\!1,n\!-\!1}\!\!\!\!\!\!\!\! & 0 \\
0 & \frac 12 V'(Q)_{nn}
\end{array}\right]  \cr
&\& + \left[
\begin{array}{cc}
\nabla_Q V'(x)_{n\!-\!1,n\!-\!1} & \nabla_Q V'(x)_{n\!-\!1,n}\\
 \nabla_Q V'(x)_{n,n\!-\!1} & \nabla_Q V'(x)_{nn}\end{array}
\right]\left[
\begin{array}{cc}
0&-\gamma_n\\
\gamma_n &0\end{array}\right] \ .
\eea
It follows from (\ref{diagVprime}) that the diagonal  $V'(Q)$ terms in $\VV_n(x)$ are 
cancelled by the sum in the last term of (\ref{VnAn}), giving the stated result
(\ref{fdd}), (\ref{formulafordd}).

Combining the differential, recursion  and deformations relations
(\ref{Pdefn}),  (\ref{Recn}), (\ref{Cdefn}), (\ref{Trsdefn}) and  (\ref{Ardefn}), 
the fact that the invertible matrices $\BG_n$ are simultaneous fundamental systems 
for all these equation implies the compatibility of the cross-derivatives; i.e., the 
corresponding set of zero-curvature equations.
\bc
\label{diffdefrec}
For $n\geq 0$ the  set of PDE's and recursion equations
\bea
\hbar \pa_x\BG_n(x)  &\&=\DD_n(x)\BG_n(x) , \qquad
\hbar \pa_{a_i}\BG_n(x)   = {\mathcal A}_{i;n}(x)\BG_n(x) \cr
\hbar \pa_{c_r}\BG_n(x)  &\&=\CC_{r;n}(x)\BG_n(x) , \qquad
\hbar \pa_{t_{r,J}} \BG_n(x)   = \TT_{r,J;n}(x)\BG_n(x) \cr
\BG_{n+1}(x)&\&=R_n(x)\BG_n(x) 
\eea
are simultaneously satisfied by the invertible matrices $\BG_n(x)$,
and hence the zero-curvature equations 
\bea
 [\hbar \pa_x-\DD_n, \ \hbar \pa_{a_i}  -  {\mathcal A}_{i;n}]&\&=0  , \quad
  [\hbar \pa_x-\DD_n, \ \hbar \pa_{a_i}  -  \CC_{r;n} ]=0  , \cr
 [\hbar \pa_x-\DD_n, \ \hbar \pa_{a_i} - \TT_{r,J;n} ]&\&=0 \ , \quad
 [\hbar \pa_{a_i}  = {\mathcal A}_{i;n}, \ \hbar \pa_{a_i}-\CC_{r;n}]=0 , \cr
  [\hbar \pa_{a_i}  - {\mathcal A}_{i;n},  \hbar \pa_{a_i}-\TT_{r,J;n}] &\&=0 \ , \quad
 [ \hbar \pa_{a_i} -\CC_{r;n},  \hbar \pa_{a_i}-\TT_{r,J;n} ]=0  , \cr
  \hbar\pa_{a_i}R_n  &\&= {\mathcal A}_{i;n+1} R_n - R_n {\mathcal A}_{i;n}  , \cr
  \hbar\pa_{c_r}R_n &\&=\CC_{r;n+1} R_n - R_n \CC_{r;n} , \cr
 \hbar\pa_{t_{r,J}}R_n  &\&= \TT_{r,J;n+1} R_n - R_n \TT_{r,J;n}\
\eea
are satisfied.
\ec


\br 
 (The Riemann--Hilbert method.)

The Riemann-Hilbert method for characterizing orthogonal polynomials 
\cite{IKF, DKMVZ} provides an alternative approach to deriving the results 
of this section. This is a well-established approach, and will not be 
developed in detail here, except to indicate briefly how it could be applied
to deducing the differential and deformation  equations satisfied by the f
undamental systems.

The fundamental system $\BG_n(x)$ has, by construction, a jump-discontinuity 
across any of the contours defining the orthogonality measure.
denoting the limiting values when approaching any of these contours from
the left or the right by$\BG){n,_\pm}$ we have the jump disconinuity
conditions
\be
\Psi_{+}(x) = \Psi_{-}(x)\left[
\begin{array}{cc}
1 & 2i\pi \varkappa_j\cr
0&1
\end{array}
\right]\ ,\qquad x\in \gamma_j
\label{jumps}
\ee
Furthermore, the local asymptotic behavior near the
singularities at $\infty$ are specified as in section \ref{Formal}. To be
more precise the function $\Gamma_n(x)$  has local formal
asymptotic form, within any of the Stokes sectors, 
\bea
\Psi(x)\sim\left\{
\begin{array}{cc}
\ds C_r\bigg(\1 + \mathcal O(x-c_r)\bigg){\rm e}^{-\frac 1 {2\hbar}T_r(x)\sigma_3} & x\to c_r\\[10pt]
\ds A_j \bigg(\1 + \mathcal O(x-a_j)\bigg) {\rm e}^{-\varkappa_j \ln(x-a_j)\sigma_+} & x\to
 a_j\\[10pt]
\ds C_0 \left(\1 + \mathcal  O \left(\frac 1 x\right)\right) {\rm e}^{-\frac 1
  {2\hbar} T_0(x)\sigma_3 + (n-\sum_{r}t_{0,r})\sigma_3\ln (x)} & x\to \infty
\end{array}
\right.
\eea

 It follows from the  usual argument based on Liouville's theorem that 
 any two fundamental solutions two (with the same Stokes matrices, 
 given in fact by the  same matrices (\ref{jumps}) ) satisfying the above 
 Riemann--Hilbert conditions are equal, within a constant scaler multiple. 
 Also, from Liouville's theorem it follows that the first column
of $\Psi(x)$  consists of polynomials (the orthogonal polynomials).
Using similar arguments one can show that the following matrix is
rational with poles, of the correct order, at the singular points
$c_r,a_j,\infty$:
\be
\DD_n(x) := \pa_x\Psi(x)\Psi(x)^{-1}  + \frac 1 2 V'(x) \1.
\ee
By comparing the local singular behavior of the logarithmic (matrix) derivatives of
any two solutions and applying Liouville's theorem,  it again that these 
globally combine to define rational matrix functions which give the deformation matrices 
with respect to the various parameters at the poles.
\er

\section{Spectral curve and spectral invariants}
\label{Spectral}

The aim of this section is to express the spectral curve of the ODE
(\ref{Pdefn}) (i.e., the characteristic equation of $\DD_n(x)$) in
terms of the partition function. In fact we will prove an exact finite $n$
analog (Thm. \ref{eval}) of the formul\ae\  that are obtained by variational methods 
in the $n\to \infty$ limit \cite{DGZ}. We start by expressing the explicit relation 
between the partition function and the spectral curve of the isomonodromic system.

\subsection{Virasoro generators and the spectral curve}

To express the result in a compact form, introduce the following {\bf local Virasoro} generators
\bea
 \mathbb V^{(r)}_{-J} &\&:= \sum_{M=1}^{d_r- J} \,M t_{r,M+ J}\, \frac
\pa{\pa t_{r,M}}\ ,\qquad  J=0,\dots , d_r-1\\
 \mathbb V^{(0)}_{- J} &\&:= \sum_{J=1}^{d_0- J} \,M t_{0,M+ J}\, \frac
\pa{\pa t_{0,M}}\ ,\qquad  J=0,\dots ,d_0 - 1 ,
\eea
in terms of which we define the following differential operator  with coefficients that are rational functions of $x$
\be
\mathbb D(x):= \sum_{i=1}^{{L}} \frac
{1}{x-a_i}\frac {\pa }{\pa a_i} 
-\sum_{ J=0}^{d_0-3} x^J\mathbb V^{(0)}_{-J-2}
-\sum_{r=1}^{K} \sum_{ J=2}^{d_r+1} \frac
1{(x-c_r)^{ J}} \mathbb V^{(r)}_{2-J}
-\sum_{r=1}^{K}\frac 1{x-c_r}\frac{\pa}{\pa c_r} \label{Dvir}\ .
\ee

\bt 
\label{spectral}
The characteristic polynomial of the matrix $\DD_n(x)$ in  the differential system (\ref{Pdefn})
 is  given by 
\beaq
\det\big(y-\DD_n(x)\big) &\&=  y^2-yV'(x) +   \hbar\left\langle \tr \frac
    {V'(M)-V'(x)}{M-x}\right\rangle  -
\sum_{i=1}^{{L}} \frac {\hbar^2}{x-a_i}\pa_{a_i}\ln(\ZZ_n)  \label{expect}\\
&\&= y^2-yV'(x) +   \hbar \tr_n
\left(\frac {V'(Q)-V'(x)}{Q-x}\right) -
\sum_{i=1}^{{L}} \frac {\hbar^2}{x-a_i}\pa_{a_i}\ln(\ZZ_n)  \label{spok}\\
&\&=y^2-yV'`(x) +
n \sum_{J=1}^{d_0 - 1} t_{0,J+1} x^{J-1} -\hbar^2 \mathbb D(x)\ln\ZZ_n
\label{curveDlnZ} ,
\eeaq
\et
and the quadratic trace invariant is
\be
\tr \DD_n(x)^2 =V'(x)^2
 -2n \sum_{J=1}^{d_0 -1 } t_{0,J+1} x^{J-1} 
+2\hbar ^2\mathbb D(x) \ln\ZZ_n. \label{trd2} 
\ee
\noindent {\bf Proof:}
The equivalence of (\ref{expect}) and (\ref{spok}) follows
from the well--known relation
\be
\left< \tr(f(M))\right>= \tr_n (f(Q))
\ee
for any scalar function $f(x)$ for which the $\left<\tr(f(M))\right>$ is a convergent integral.
The equivalence of (\ref{curveDlnZ}) and (\ref{trd2}) follows from (\ref{trd1}).

To prove (\ref{spok}) we use the recursion relation  (\ref{Recn}) and the explicit expression of $\DD_n$, we obtain
\bea
\DD_{n+1}&\&=R_n\DD_n {R_n}^{-1} + \hbar  R'_n {R_n}^{-1}\\
R'_n{R_n}^{-1}&\&=\left[\begin{array}{cc} 0& 0\\ \frac 1{\gamma_{n+1}}
    & 0 
\end{array}
\right]\ ,\qquad 
{R_n}^{-1}R_n' = \left[\begin{array}{cc} 0 &-\frac 1 {\gamma_n}\\
    0& 0 \end{array}\right] \ .
\eea
Therefore,
\bea
 {\rm Tr}\left(\DD_{n+1}(x)^2\right)&\&= {\rm Tr}\left(\DD_n(x)^2\right) + 2
\hbar {\rm
  Tr}\left( \DD_n(x) {R_n}^{-1} R'_n\right) +\hbar^2 {\rm Tr}\left(\left(
R'_nR_n^{-1}\right)^2\right)   \\
&\&=  {\rm Tr}\left(\DD_n(x)^2\right)  -2 \hbar \left(\frac {V'(Q)-V'(x)}{Q-x}
\right)_{nn} - 2\hbar^2 \sum_{i=1}^{{L}} \frac {\varkappa_{i}\psi_n^2(a_i)}{x-a_i} 
\label{tr2psi} \\
&\&=   {\rm Tr}\left(\DD_n(x)^2\right)  - 2 \hbar \left(\frac {V'(Q)-V'(x)}{Q-x}
\right)_{nn} +2\hbar^2 \sum_{i=1}^{{L}} \frac {1}{x-a_i}\pa_{a_i}\ln(h_n),
\label{tr2lnh}
\eea
where we have used eqs.~(\ref{fdd}), (\ref{formulafordd}) in (\ref{tr2psi}) and (\ref{Ainnlnpsi}) in (\ref{tr2lnh}).
These equations imply that 
\be
{\rm Tr}(\DD_{n}(x)^2) ={\rm Tr}(\DD_{1}(x)^2) - 2\hbar \sum_{j=1}^{n-1}
\left(\frac {V'(Q)-V'(x)}{Q-x}\right)_{jj} +2\hbar^2 \sum_{j=1}^{n-1}
\sum_{i=1}^{{L}} \frac {1}{x-a_i}\pa_{a_i}\ln(h_{j}).
\label{ads}
\ee
From the definition of $\DD_1$, we have
\be 
\DD_1 = \hbar \left[\begin{array}{cc} \pi_0' & \phi_0' \\
  \pi_1' & \phi_1' \end{array}\right]  \left[\begin{array}{cc} \pi_0& \phi_0 \\
  \pi_1 & \phi_1\end{array}\right]^{-1}.
  \ee
  Using (\ref{CDdet}), this gives
  \bea
  \det (\DD_1(x)) &\&=\hbar^2 \gamma_1 e^{-{1\over \hbar} V(x)} \phi_0' \pi_1'  \\
  &\&= \hbar^2 \sqrt{h_1\over h_0} e^{-{1\over \hbar} V(x)}  \sqrt{1\over h_0 h_1} 
  \left[e^{+{1\over \hbar} V(x)} \int _\varkappa {e^{-{1\over \hbar} V(z)}  \over x-z} dz\right]' \\
  &\&= {\hbar^2\over h_0} \left[ {1\over \hbar} V'(x)  \int _\varkappa {e^{-{1\over \hbar} V(z)}  \over x-z} dz 
  -  \int _\kappa {e^{-{1\over \hbar} V(z)}  \over (x-z)^2} dz  \right] \\
     &\&= {\hbar^2\over h_0} \left[ {1\over \hbar} V'(x) \int
    _\varkappa {e^{-{1\over \hbar} V(z)}  \over x-z} dz  -  \int
    _\varkappa e^{-{1\over \hbar} V(z)} {\pa \over \pa  z}
    \left({1\over x-z} \right)dz   
\right] \\
     &\&=  \left[ \hbar   \int _\kappa  {(V'(x) -V'(z))\psi_0^2(z) \over x-z} dz 
        + \hbar^2 \sum_{i=1}^K {\kappa_i \psi_0^2(a_i) \over x-a_i}   \right]  , 
    \eea
    and hence
    \bea 
    \tr (\DD_1^2(x)) &\&= - 2 \det(\DD_1(x)) + \tr(\DD_1(x))^2\\
  &\&= ( (V'(x))^2 -2 \hbar   \int _\kappa  {(V'(x) -V'(z))\psi_0^2(z) \over x-z} dz 
 -2 \hbar^2 \sum_{i=1}^K {\kappa_i \psi_0^2(a_i) \over x-a_i}\\
  &\&= (V'(x))^2 - 2\hbar\left(\frac {V'(x)-V'(Q)}{x-Q}\right)_{00}+ 2\hbar^2\sum_{i=1}^{{L}} ,
\frac {1}{x-a_i} \pa_{a_i}\ln(h_0) \label{asd}. 
\eea
Combining this  with (\ref{ads}) gives
\be
{\rm Tr}(\DD_{n}(x)^2) = (V'(x))^2  - 2\hbar \sum_{j=1}^n
\left(\frac {V'(Q)-V'(x)}{Q-x}\right)_{jj} +2\hbar^2 \sum_{j=1}^n
\sum_{i=1}^{{L}} \frac {1}{x-a_i}\pa_{a_i}\ln(h_{j}),
\label{Trdn2}
\ee
which, taking the expression (\ref{partit}) for the partition function into 
account,  completes the proof of eq.~(\ref{spok}). 

We now proceed to the proof of  eq.~(\ref{curveDlnZ}).
By expanding the third term on the right  of (\ref{spok}),  we obtain
\bea  
&\& \frac {V'(x)-V'(Q)}{x-Q}= \sum_{J=1}^{d_0-1}
t_{0,J+1}\sum_{M=0}^{J-1} x^M Q^{J-M-1} + \sum_{r=1}^{K}
\sum_{J=1}^{d_r+1} t_{r,J-1}\sum_{M =0}^{J-1}\frac
1{(x-c_r)^{J-M}} (Q-c_r)^{-M-1}  \cr
&\& = 
\sum_{J=1}^{d_0-1} t_{0,J+1} x^{J-1}\1 +
\sum_{J=2}^{d_0-1}\sum_{M=0}^{J-2} x^M t_{0,J+1} Q^{J-M-1}  
 + \sum_{r=1}^K\sum_{J=1}^{d_r+1} t_{r,J-1} \sum_{M=0}^{J-1} \frac
1{(x-c_r)^{J-M}} (Q-c_r)^{-M-1} \cr
&\& = 
\sum_{J=1}^{d_0-1} t_{0,J+1} x^{J-1}\1 +
\sum_{M=0}^{d_0-3} x^M \sum_{J=M+2} ^{d_r}
t_{0,J} Q^{J-M-1} + 
\sum_{r=1}^{K} \sum_{M=1}^{d_r+1} \frac
1{(x-c_r)^{M}}\sum_{J=M}^{d_r+1} t_{r,J-1} (Q-c_r)^{M-J-1}\cr
&& \label{VQ}
\eea
Now recall that for any deformation matrix $X$ corresponding to an
infinitesimal variation $\pa$ we have
\be
\sum_{j=0}^{n-1}X_{jj} = -\frac \hbar 2 \pa\ln(\ZZ_n) . \label{tracedef}
\ee
Summing the diagonal terms of (\ref{VQ}) up to $n-1$
and substituting into (\ref{spok}) we therefore obtain
\beaq
 {\rm Tr}(\DD_{n}(x)^2)&\&= (V'(x))^2 
-2n\hbar \sum_{J=1}^{d_0-1} t_{0,J+1} x^{J-1}  \cr
&&{\hskip -10 pt}
-2\hbar^2\sum_{M=1}^{d_0-2} x^M \sum_{J=M+1} ^{d_0}
 (J-M-1) t_{0,J+1}  \frac{\pa\ln(\ZZ_n)}{\pa{t_{0,J-M -1}}} \cr
&&{\hskip -10 pt}
-2\hbar^2 \sum_{r=1}^{K} \sum_{M=2}^{d_r+1} \frac
1{(x-c_r)^{M}}\sum_{J=M}^{d_r+1} (J-M +1) t_{r,J-1}
\frac{\pa\ln(\ZZ_n)}{\pa{t_{r,J-M +1}}}   \cr
&&{\hskip -10 pt}
-2\hbar^2\sum_{r=1}^{K}\frac 1{x-c_r}\frac{\pa\ln(\ZZ_n)}{\pa c_r}
+2\hbar^2\sum_{j=0}^{n-1} \sum_{i=1}^{{L}} \frac {1}{x-a_i}\frac
{\pa\ln(\mathcal  Z_n)}{\pa a_i}
\label{TrD2difflnZ}
\eeaq
Using eq.~(\ref{tracedef}) we finally get 
\beaq
\hspace{-1cm}\det\big(y-\DD_n(x)\big)
&\!\!\!\!\!\!=&\hspace{-8pt}  y^2-yV'(x) +
n \hbar\sum_{J=1}^{d_0-1} t_{0,J+1} x^{J-1} +
\hbar^2 \sum_{M=0}^{d_0-3} x^M \sum_{J=M+2} ^{d_0 -1}
(J-M-1) t_{0,J+1} \frac{\pa\ln(\ZZ_n)}{\pa{t_{0,J-M-1}}} \cr
&&{\hskip -10 pt}+ 
\hbar^2\sum_{r=1}^{K} \sum_{M=2}^{d_r+1} \frac
1{(x-c_r)^{M}}\sum_{J=M}^{d_r+1} (J-M +1)t_{r,J-1}
\frac{\pa\ln(\ZZ_n)}{\pa{t_{r,J-M +1}}}   \cr
&&{\hskip -10 pt}+ 
\hbar^2\sum_{r=1}^{K}\frac 1{x-c_r}\frac{\pa\ln(\ZZ_n)}{\pa c_r} -
\hbar^2\sum_{i=1}^{{L}} \frac {1}{x-a_i}\frac{\pa\ln(\ZZ_n)}{\pa{a_i}} ,
\eeaq
which completes the proof of eq.~(\ref{curveDlnZ}).

\subsection{Spectral residue formul\ae}

  Theorem \ref{spectral}, which determines all the coefficients of the spectral
   curve as logarithmic derivatives of the partition function, may be
   expressed in another form, in which the individual deformation
   parameters, as well as the logarithmic derivatives with respect to
   them, may be directly expressed as spectral invariants. 
The characteristic equation of $\DD_n(x)$ 
 \be
\det\left(y (x) {\bf I}-\DD_n(x)\right) =0 , \label{spectralcurven}
\ee
defines a hyperelliptic curve $\CC_n$ as a $2$--sheeted branched cover of the 
Riemann sphere, on which $y$ is a meromorphic function. It follows
   from  (\ref{fdd}) and Theorem \ref{spectral} that $y$, viewed as a
   double valued function of $x$, has the same pole structure and
   degree as $\DD_n(x)$ at the points $\{c_0=\infty, c_r, a_i\}$, but
   that the points $\{a_i\}$ are branch points. 
   
Let $Y_{\pm}(x)$ denoted the two values of $y(x)$. Defining
\be
W(x) :=\hbar \tr_n\frac {V'(x)-V'(Q)}{x-Q}  -\sum_{j=1}^{L}\frac
{\hbar^2\pa_{a_j}\ln\ZZ_n}{x-a_j},
\ee
it follows from the explicit expression  (\ref{spok}) for the spectral curve 
that, near any of the poles  $c_0=\infty,\ c_1,\dots , c_K$, the two branches have 
the asymptotic form
\beaq
 Y_\pm(x)&\& =  \frac 12 V'(x) \pm\sqrt {\frac 1 4 (V'(x))^2 - W } \cr
&\&\sim \left\{1\atop 0\right\} V'(x) \mp\frac 1{V'(x)}
\left(W + \frac {W^2}{(V(x)')^2} + \dots\right) + \left\{\begin{array}{ll}
\mathcal O(x^{-2d_0-1}) & x\to\infty\\
\mathcal O((x-c_r)^{2 d_r+5}) & x\to c_r
\end{array}\right.  \label{Ypm}
\eeaq

\bt
\label{eval}
The following residue formul\ae\   express  the deformation
parameters and the logarithmic derivatives of $\ZZ_n$
as spectral invariants of the matrix $\DD_n(x)$.
\beaq
t_{0,J} &\&=\frac 1 {2i\pi} \oint_{\infty} \frac {Y_+(x)}{x^J}dx,  \quad   J=1\dots d_0 \\
t_{r,J} &\&=  \frac 1 {2i\pi} \oint_{c_r} (x-c_r)^{J}Y_+(x)dx,  \quad r=1,\dots,K,\ J=1,\dots, d_r\label{coords} \\
  \hbar^2 \pa_{t_{0,0}}\ln\ZZ_n &\&=- \frac 1 {2i\pi}\oint_\infty
Y_-(x)dx\label{trivia}  = -n \hbar ,   \label{t00}\\ 
\hbar^2 \pa_{t_{0,J}}\ln \ZZ_n &\&=  - \frac 1 {2 i J\pi}
\oint_{\infty} Y_-(x) x^J dx,
  \quad J=1, \dots , d_0 \label{inf}\\
 \hbar^2  \pa_{t_{r,J}}\ln \ZZ_n &\&= - \frac 1 {2 i J\pi} \oint_{c_r} Y_-(x)
\frac 1{(x-c_r)^J} dx,  \quad r=1,\dots,K,\ J=1,\dots,d_r   \label{trJ}\\
\hbar^2  \pa_{c_r}\ln \ZZ_n &\&= -\frac 1 {2i\pi} \oint_{c_r} Y_-(x)
T'_r(x) dx\label{cr}  ,  \quad  r=1,\dots,K,   \\
 \hbar^2  \pa_{a_j}\ln \ZZ_n &\&=\frac 1 {4\pi i} \oint_{a_j}
\tr(\DD_n^2(x))dx \label{aj}\ .
\eeaq
\et

{\bf Proof.}  The second equality in eq.~(\ref{trivia}) follows from the fact that
$t_{0,0}$  appears in the integral (\ref{partit}) defining $\ZZ_n$ only  in the
 overall normalization factor $e^{-{t_{0,0}\over \hbar}}$.
The proof of the other relations is based on formula (\ref{TrD2difflnZ}), 
which was proved  in the demonstration of Theorem \ref{spectral}. 
From this formula  it follows that, for all deformations except the ones with
 respect to $t_{0,d_0}, t_{0,d_0 - 1},a_j$,
\be
\hbar^2 \pa\ln\ZZ_n =\frac 1 2  \res{} \frac{\pa V(x)}{V'(x)} \tr (\DD_n^2(x)) ,
\ee
where $\pa V(x)$ denotes the deformation $\pa$ of the 
potential $V$, and the residues are taken at the corresponding
singularity $c_r$. Considering the various deformations associated to
the poles and end-points we have: 
 \newline \noindent
{\bf At infinity}
\bea
\hbar^2 \pa_{t_{0,J}} \ln\ZZ_n  &\&=\frac 1 2  \res{x=\infty} \frac{x^J/J}{V'(x)}
\tr (\DD_n^2(x))  \cr
&\&=\frac 1 2   \res{x=\infty} \frac{x^J/J}{V'(x)}\left( (V'(x))^2 -2
\hbar \tr_n \frac {V'(x)-V'(Q)}{x-Q} +2\sum \frac {\hbar^2}{x-a_j}\pa_{a_j} \ln
\ZZ_n\right) \cr
&\&=- \hbar \res{x=\infty} \frac{x^J/J}{V'(x)}\tr_n \frac {V'(x)-V'(Q)}{x-Q}
= \frac \hbar  J \tr_n Q^{J} \ ,\ \ J=1,\dots d_0-2.\label{inf1}
\eea
Note that this computation does not provide the derivatives with respect to the two highest
coefficients $t_{0,d_0}$ and $t_{0,d_0 -1}$, which will be computed
below.
Moreover we should remark that the last equality follows from the
following interchange of order of integrals 
\bea
 \res{x=\infty} \frac{x^J/J}{V'(x)}\tr_n \frac {V'(x)-V'(Q)}{x-Q}  =
 \sum_{j=0}^{n-1}\res{x=\infty} \frac{x^J/J}{V'(x)}\int_{\varkappa}
 \frac {V'(x)-V'(z)}{x-z}{\pi_j}^2(z){\rm e}^{-\frac 1 \hbar V(z)}{\rm
   d}z =\cr 
= \sum_{j=0}^{n-1}\int_{\varkappa}
 \res{x=\infty} \frac{x^J/J}{V'(x)}\frac {V'(x)-V'(z)}{x-z}{\pi_j}^2(z){\rm e}^{-\frac 1 \hbar V(z)}{\rm
   d}z = \sum_{j=0}^{n-1}\int_{\varkappa}
 \res{x=\infty} \frac{z^J}J{\pi_j}^2(z){\rm e}^{-\frac 1 \hbar V(z)}{\rm
   d}z=\frac 1 J \tr_n Q^J\ . 
\eea
The exchange is justified by the usual arguments observing that the
expression $\frac {V'(x)-V'(z)}{x-z}$ has no singularities at
coinciding points $x=z$ (away from the singularities of $V'$).

{\bf At the poles $c_r$}
\bea
\hbar^2 \pa_{t_{r,J}} \ln\ZZ_n  &\&=\frac 1 2\res{x=c_r} \frac{(x-c_r)^{-J}/J}{V'(x)}
\tr (\DD_n^2(x))  \cr
&\&=- \res{x=c_r} \frac{(x-c_r)^{-J}/J}{V'(x)}\hbar \tr_n \frac {V'(x)-V'(Q)}{x-Q}
= \frac \hbar J \tr_n (Q-c_r)^{-J}\ ,\ \ J=1,\dots, d_r;\cr
\hbar^2 \pa_{c_r} \ln\ZZ_n  &\&=\frac 1 2\res{x=c_r} \frac{T'_r(x)}{V'(x)}
\tr (\DD_n^2(x))  \cr
&\&=- \hbar \res{x=c_r} \frac{T'_r(x) }{V'(x)}\tr_n \frac {V'(x)-V'(Q)}{x-Q}
= \frac  \hbar J \tr_n T'_r(Q)\ .\label{cr1}
\eea
The last equalities in (\ref{cr1}) are obtained by a similar
argument used for the deformations at $c_0=\infty$ here above.

{\bf At the endpoints $a_j$}
\bea
\hbar^2 \pa_{a_j}  \ln\ZZ_n  &\&=\frac 1 2\res{x=a_j} 
\tr (\DD_n^2(x))  \cr
&\&
=\frac 1 2 \res{x=a_j}\left( (V'(x))^2 -2
\hbar\tr_n \frac {V'(x)-V'(Q)}{x-Q} +2\sum_{j=1}^{L}\frac {\hbar^2 }{x-a_j}\pa_{a_j} \ln
\ZZ_n\right)  \cr
&\&=\hbar^2  \pa_{a_j} \ln
\ZZ_n\ .\label{aj1}
\eea

The relations (\ref{coords}) now follow immediately from (\ref{Ypm}). 
The determination  $Y_-$  has the  asymptotic behavior
\be
Y_-(x)\sim \frac 1{V'(x)}\left(\hbar 
\tr_n\frac {V'(x)-V'(Q)}{x-Q} -\sum_{j=1}^{L}\frac
{\hbar^2} {x-a_j}\pa_{a_j}\ln\ZZ_n -\frac {n^2\hbar^2}{x^2} \right)+ \left\{\begin{array}{ll}
\mathcal O(x^{-2 d_0-1}) & x\to\infty\\
\mathcal O((x-c_r)^{2 d_r+5}) & x\to c_r
\end{array}
\right.
\label{asymptY}
\ee

Identities (\ref{t00}),  (\ref{trJ}) and (\ref{cr}) follow immediately from the
expressions in (\ref{inf1}),  (\ref{cr1}) and the asymptotic forms (\ref{asymptY}),
as do the identities (\ref{inf}) for $J\le d_0 -2$.
For the remaining two values of $J$  $(d_0 -1 ,d_0)$,  we compute
\bea
-\res{x=\infty} T_0'(x) Y_-(x)dx 
&\&=-\res{x=\infty} (V'(x)+\mathcal O(1/x)) \frac 1{V'(x)}\left( W - \frac
{n^2\hbar^2}{x^2} + \mathcal O\left(x^{-3})\right)\right)dx  \cr
&\&=-\sum_{r=1}^K \hbar \tr_n T'_r(Q)
+\sum_{j=1}^{L} \hbar^2 \pa_{a_j} \ln\ZZ_n {=}\hbar  \tr_n T'_0(Q) ,
\eea
where the last  equality  follows from eq.~\ref{diagVprime} (translational invariance).
This identity, together with eqs.~(\ref{inf}) for $j\le d_0 - 2$ implies
\be
-\res{x=\infty} x^{d_0 - 1}Y_-(x)dx =\hbar^2 \frac {\pa}{\pa t_{0,d_0 -1}} \ln \ZZ_n\ ,
\ee
which is the case $J=d_0 - 1$ of  (\ref{inf}). 
Similarly
\bea
- \res{x=\infty} xT_0'(x) Y_-(x)dx &\&= 
-\res{x=\infty} (xV'(x)+\sum_{r=1}^K t_{r,0}+\mathcal O(1/x)) \frac 1{V'(x)}\left( W - \frac
{n^2\hbar^2}{x^2} + \mathcal O\left(x^{-3})\right)\right)dx  \cr
&\&=n^2\hbar^2 -n\hbar \sum_{r=1}^K t_{r,0}  - \sum_{r=1}^K\hbar \tr_n QT'_r(Q)
+\sum_{j=1}^{L} \hbar^2 a_j\pa_{a_j} \ln\ZZ_n \cr 
&\& = \hbar \tr_n QT'_0(Q) 
\eea
where the last equality holds because of dilation invariance. This,
together with the above proves (\ref{inf})  for $J=d_0$.

The last formula (\ref{aj}) follows from equating the residues at
the poles  $x=a_j$ in eq.~(\ref{trd2}).

\br
In the formul\ae\ (\ref{trivia}), (\ref{inf}), (\ref{cr}), (\ref{aj}) we may
replace $Y_-$ by $-(Y_+ -V'(x))$; this corresponds to the fact that, in the 
large $n$ limit,  the behavior of $Y(x)$ on the {\em physical sheet} 
(i.e. $Y_+$) is related to the resolvent of the model by 
  \be
  Y_+ = V'(x) + \hbar \left\langle\tr (x-M)^{-1}\right\rangle.
  \ee
\er

\section{Isomonodromic Tau function}
\label{Isomonodromic}

\subsection {Isomonodromic deformations and residue formula}
\label{residuetau}

In this section we briefly recall the definition of the isomonodromic
tau-function given  in  \cite{JMU} and compute its logarithmic derivatives in 
the present  case in order to compare it with the partition function. This will lead to
the main result of this section, Theorem \ref{main}, which explicitly gives this relation.

Consider a  rational covariant derivative operator on a rank $p$ vector bundle over $\Cbb P^1$ 
\be
\DD_x = \pa_x - {\mathcal A}(x) \ ,
\ee
where the connection component ${\mathcal A}(x)$ is a $p \times  p$ matrix,
rational in $x$. Deformations of such an operator that preserve its (generalized) monodromy  
 (i.e. including the Stokes' data) are determined infinitesimally by requiring 
 compatibility of the equations
\bea
 \pa_x\Psi(x) &\&= {\mathcal A} \Psi(x) \\
 \pa_{u_i} \Psi(x) &\&= \mathcal U_i(x) \Psi(x)\ ,  \quad  i = 1, \dots  \ .
\eea
where in the second set of equations $U_i (x)$ are also  $p \times  p$ matrices,
rational in $x$, viewed as components of a connection over the extended space
consisting to the product of $\Cbb P^1$ with the space of deformation
 parameters $\{u_1, \dots \}$.
The invariance of the generalized monodromy of $\mathcal \DD_x$
follows \cite{JMU} from the compatibility of this overdetermined system,
which is equivalent to the zero-curvature equations
\be
[\pa_x - {\mathcal A}(x),\pa_{u_i} - \mathcal U_i(x)]=0\  , \quad 
[\pa_{u_i} - \mathcal U_i(x),\pa_{u_j} - \mathcal U_j(x)]=0\  
\ee
Near a pole $x=c_\nu$ of ${\mathcal A}(x)$ a fundamental solution can be found that
has the {\em formal} asymptotic behavior,  in a suitable sector:
\be
\Psi(x)\sim C_\nu  Y_\nu(x){\rm e}^{T_\nu(x)}
\ee
where $C_\nu$ is a constant matrix, 
\be
Y_\nu(x)=\1 + \mathcal O(x-c)
\ee
 is a formal power series in the local parameter $(x-c_\nu)$ (or $1/x$ for
 the pole at infinity) and $T_\nu (x)$ is a Laurent-polynomial matrix in the local parameter,
plus a possible logarithmic term $t_0 \ln(x-c)$. In the generic case $T_\nu(x)$ is a
diagonal matrix, and, more generally, may be an element of a maximal
Abelian subalgebra containing an element with no multiple eigenvalues.
  The locations of the pols $c_\nu$ and the  coefficients of the nonlogarithmic 
  part of $T_\nu(x)$  are the independent deformation parameters.
 The deformation  of the connection matrix $A(x)$ is determined by the
requirement that the (generalized) monodromy data be independent of all
these isomondromic deformation parameters.

Given a solution of such an isomonodromic deformation problem, one is led to 
consider the  associated isomonodromic $\tau$-function \cite{JMU},
determined by integrating the following closed differential on the space of
deformation parameters 
\be
\omega : = \sum_\nu\res{x=c_\nu}\tr
\left(Y_\nu^{-1}Y_\nu'\cdot dT_\nu(x)\right)  =  d\ln \tau^{IM}\ ,\label{closed}
\ee
where the sum is over all poles of ${\mathcal A}(x)$ (including possibly one
at $x=\infty$), and the differential is over all the independent isomonodromic 
deformation parameters , In the  present situation ${\mathcal A}(x)$ is 
our $2\times 2$ matrix  $\DD_n(x)$ and the (generalized) monodromy of the operator
$\pa_x-\DD_n(x)$ is invariant under changes in the parameters $c_r,\
t_{r,J}\ ,a_j$ and $n$.

\subsection{Traceless gauge}
\label{Formal}

For convenience in the computations we perform a scalar gauge 
transformation of the ODE by choosing {\em quasipolynomials} rather
than polynomials. Explicitly we 
 set 
\bea
{\Psi_n} (x)&\&:= {\rm e}^{-\frac 1 {2\hbar} V(x)} \BG_n(x) =
\left[\begin{array}{cc}
\psi_{n-1}(x) & \wt\psi_{n-1}(x)\cr
\psi_n(x) & \wt \psi_n(x)
\end{array}
\right]\cr 
\hbar \Psi_n'(x) &\&= {\mathcal A}_n(x) \Psi_n(x)\cr
{\mathcal A}_n(x) &\&= \DD_n(x) - \frac 1 2 V'(x)\1 \label{newgauge} ,
\eea
where
\be
 \wt\psi_n:= e^{-{1\over 2\hbar} V(x)} \psi_n
 \ee
In this gauge the matrix of the ODE is traceless and the infinitesimal deformation
matrices  are transformed correspondingly by addition of  the identity element 
multiplied by the derivatives of $-{1\over 2 \hbar} V(x)$ with respect to the
 parameters $\{c_r, t_{rJ}, a_j\}$.
This choice gives a consistent  reduction of the general  $\grgl(p, {\bf C})$ 
isomonodromic deformation problem  to $\grsl(p, {\bf C})$. (To be precise, this
would require a further, $x$--independent diagonal gauge transformation of the form 
$\diag(h_{n-1}^{-1\over 2} , h_n^{1\over 2})$ to render the infinitesimal  deformation 
matrices also traceless.)

 At each of the poles $c_0:=\infty,c_1,\dots$ we then 
have the following asymptotic expansions. (To simplfy notation, the index $n$
is omitted in labelling  the fundamental system $\Psi$ and its local asymptotic form .)
\be
\Psi(x)\sim C_r Y_r(x)\ \exp\left[{\left(-\frac 1 {2\hbar} T_r(x)+
 \delta_{r0} \left(n + \frac 1 {2\hbar}\sum_{r\geq 1} t_{r,0}\right)\ln(x)\right)  \sigma_3}
 \right ]\label{Gammacrasympt}\ .
\ee
Here we have set 
\bea
 Y_0(x)&\&:= \1 + \sum_{k=1}^\infty \frac {Y_{0;k}}{x^k}\ ,\qquad 
C_0 = \left[\begin{array}{cc}
0 & \sqrt{h_{n\!-\!1}} \\ \frac 1{\sqrt{h_{n}}} &0
 \end{array}\right] \cr
 Y_r(x)&\& := \1 + \sum_{k=1}^\infty Y_{r;k}(x-c_r)^k\ ;\qquad 
C_r =  \left[\begin{array}{cc}
\pi_{n\!-\!1}(c_r) {\rm e}^{-\frac {\check{V}_r(c_r)}{2\hbar}}  
& (c_r-Q)^{-1}_{n\!-\!1,0}\sqrt{h_0} e^{{\check{V}_r(c_r)\over2 \hbar}}, \\
\pi_{n}(c_r) {\rm e}^{-\frac {\check{V}_r(c_r)}{2\hbar}}  &
(c_r-Q)^{-1}_{n,0}\sqrt{h_0}e^{{\check{V}_r(c_r)\over2 \hbar}}  \\
 \end{array}\right]
 \label{Trexp}, 
\eea
where $\check{V}_r(x) = V(x) - T_r(x)$ is the holomorphic part of the potential at $c_r$.
The asymptotic forms given by (\ref{Gammacrasympt})--(\ref{Trexp})
follow from the fact that, in any Stokes' sector near $c_r$,  the
second-kind solutions behave like 
\bea
\wt\psi_n(x)&\&  \m{\sim}_{x\to \infty} {\rm e}^{\frac 1{2\hbar} V(x)} \sum_{k=n+1}^{\infty}x^{-k}
\int_\varkappa \pi_n(z){\rm e}^{-\frac 1\hbar V}z^{k-1}  = {\rm
  e}^{\frac 1{2\hbar} V(x)}x^{-n-1}\sqrt{h_n}\left(1+\mathcal O\left(\frac 1 x\right)\right)\label{psinf}\\
\wt\psi_n(x)&\& \m{\sim}_{x\to c_r} {\rm e}^{\frac 1{2\hbar}V(x)}\int_{\varkappa}{\rm
  d}z\frac {{\rm e}^{-\frac 1 \hbar V(z)}\pi_n(z)}{c_r-z}(1+\mathcal O(x-c_r)) \cr
  &\& = {\rm
  e}^{\frac 1 \hbar V(x)}(c_r-Q)^{-1}_{n,0}\sqrt{h_0} (1+\mathcal O(x-c_r))
\eea
Near the endpoints $a_j$ we have
\bea
\Psi(x)\sim A_j \cdot Y_j(x)\cdot \exp\left[
-\varkappa_j\ln(x-a_j)\sigma_+ \right ]\ ,\qquad \sigma_+:=\pmatrix {0&1\cr 0&0}
\cr
A_j =  \left[\begin{array}{cc}
\pi_{n\!-\!1}(a_j){\rm e}^{-\frac { V(a_j)} {2\hbar} } & {\rm
  e}^{\frac { V(a_j)} {2\hbar}}\int_\varkappa dz \frac {{\rm 
    e}^{-\frac 1 \hbar V(z)} (\pi_{n\!-\!1}(z)-\pi_{n\!-\!1}(a_j))}{a_j-z} \\
\pi_{n}(a_j){\rm e}^{-\frac { V(a_j)} {2\hbar} } & {\rm e}^{\frac { V(a_j)} {2\hbar} }\int_\varkappa dz \frac {{\rm
    e}^{-\frac 1 \hbar V(z)} (\pi_{n}(z)-\pi_{n}(a_j))}{a_j-z} ,
 \end{array}\right]\label{he}
\eea
since the matrix 
\be
\Psi(x) \cdot \exp\left(-\sigma_+\int_\varkappa dz \frac {{\rm
    e}^{-\frac 1\hbar V(z)}}{x-z} \right) = \left[
\begin{array}{cc}
\pi_{n\!-\!1}(x) {\rm e}^{- \frac {V(x)}{2\hbar} } & {\rm e}^{ \frac {V(x)}{2\hbar} }\int_\varkappa dz \frac {{\rm
    e}^{-\frac 1\hbar V(z)} (\pi_{n\!-\!1}(z)-\pi_{n\!-\!1}(x))}{x-z} \\
\pi_{n}(x){\rm e}^{- \frac {V(x)}{2\hbar}}  & {\rm e}^{ \frac {V(x)}{2\hbar} }\int_\varkappa dz \frac {{\rm
    e}^{-\frac 1\hbar V(z)} (\pi_{n}(z)-\pi_{n}(x))}{x-z}
\end{array}
\right]
\ee
is {\em analytic} in a neighborhood of $a_j$ and has the limiting
value  indicated in (\ref{he}). The function $ -\int_\varkappa {\rm
  d}z \frac {{\rm   e}^{-\frac 1\hbar V(z)}}{x-z} $ in the exponential
of the  second matrix in this formula has the same  singularity as
$\varkappa_j\ln(x-a_j)$. (The signs in (\ref{he} follows from  the orientation of 
the contour originating at $a_j$).)

The differential (\ref{closed}) can now be written 
\be
\hbar d \ln\tau_n^{IM} = -\frac 1 2 \sum_{r=0} \res{x=c_r}dT_r(x) \tr\left(
Y_r^{-1}Y_r' \sigma_3
\right) + 
\sum_j \res{x=a_j} \frac {\varkappa_j da_j}{x-a_j} \tr\left(Y_j^{-1}Y_j' \sigma_+\right)  
\ee
where the differential involves the isomonodromic parameters only
\be
d := \sum_{r=0}\left( \sum_{J=1}^{d_r} dt_{r,J}\frac \pa{\pa t_{r,J}} 
+ dc_r \frac \pa{\pa c_r} \right) + \sum_j da_j\frac
\pa{\pa a_j} = 
\sum_{r=0}^Kd_{(r)} +  \sum_j da_j\frac
\pa{\pa a_j}
\ee

We  now derive residue formul\ae\  for the deformation parameters
 and the logarithmic derivatives of the tau function for our
 rational $2\times 2$ isomonodromic deformation problem.  These
essentially are the same as the formul\ae\   of Thm. \ref{eval} giving
the latter quantities in terms of logarithmic derivatives of the
partition function of the matrix model\footnote{
The case of an arbitrary rank rational, 
nonresonant isomonodromic deformation problem will be developed elsewhere
\cite{BHHP}, together with further properties that allows us to
view these as nonautonomous Hamiltonian systems, in which the 
logarithmic derivatives of the $\tau$-function computed below, are
interpreted as the Hamiltonians generating the deformation dynamics.}.

Consider the quadratic spectral invariant near any of the
singularities: by virtue of the asymptotics (\ref{Gammacrasympt},
\ref{he}) we have, near $c_r$ and $a_j$ respectively (setting $S:=
2n\hbar + \sum_{r=1}^K t_{r,0}$) 
\bea
\tr({\mathcal A}^2(x))&\&= \hbar^2\tr((\Psi'\Psi^{-1})^2)=\hbar^2 \tr\left(\left( Y_r^{-1}Y_r'\right)^2\right) +
2\hbar \tr\left(Y_r^{-1} Y_r' \sigma_3 \left(T_r' -\frac {\delta_{r0}S} x
\right)  \right)  \cr
&&{\hskip -10 pt}+\frac 1 2
\left (T_r' -\frac {\delta_{r0}S} x  \right)^2\\
\tr({\mathcal A}^2(x))&\&=\hbar^2\tr((\Psi'\Psi^{-1})^2)= \hbar^2 \tr\left(\left(
Y_j^{-1}Y_j'\right)^2\right) + \frac {2\hbar\varkappa_j}{x-a_j}
\tr\left(Y_j^{-1} Y_j' \sigma_+\right)\ .
\eea
Taking the principal part at each singularity and using Liouville's
theorem (since $\tr{\mathcal A}^2$ is {\em a priori} a rational
function) we find
\bea
\tr({\mathcal A}^2(x)) &\&= \sum_{r=0}^K\left[ \frac 1 2 \left(T'_r-\frac
  {\delta_{r0}S} x\right )^2  +  \frac 2
  \hbar \bigg(\left(T'_r -\frac {\delta_{r0}S} x \right)\tr(Y_r^{-1}Y_r'
\sigma_3 )\bigg)\right]_{r,+}  \cr
&&{\hskip -10 pt}+ \frac 2 \hbar \sum_{j=1}^{L} \left[
\frac {\varkappa_j \tr(Y_j'(a_j)\sigma_+)}{x-a_j}
\right]
\eea
where the subscripts $_{r,+}$ mean the singular part at the pole
$x=c_r$ (including the constant for $x=c_0=\infty$).
Consider now the spectral curve of the connection $\hbar\pa_x-{\mathcal A}(x)$
\be
w^2 = \frac 1 2 \tr {\mathcal A}^2(x)\ ,
\ee
one immediately finds 
\be
w_{\pm}(x) = \pm \sqrt{ \frac 1 4
  \sum_{r}{\left(T'_r-\frac{\delta_{r0}S} x \right)^2}_{r,+}  +  \frac 1
      {2\hbar} \bigg(\left(T'_r-\frac{\delta_{r0}S} x \right)  \tr(Y_r^{-1}Y_r'
\sigma_3 )\bigg)_{r,+} + \frac 1 \hbar \sum_j \left[
\frac {\varkappa_j \tr(Y_j'(a_j)\sigma_+)}{x-a_j}
\right]}
\ee
Near any of the poles one has the asymptotic behavior
\bea
\pm w_\pm  = \left\{ 
\begin{array}{ll}
\ds \frac 1 2 T'_r - \frac {\delta_{r0}S}{2x}+ \frac 1{2\hbar T_r'(x)}  \bigg(T'_r \tr(Y_r^{-1}Y_r'
\sigma_3 )\bigg)_{r,+} +\left\{\begin{array}{ll}
\ds \mathcal O ((x-c_r)^{d_r+1}) & \hbox{ near } x=c_r\\[10pt]
 \ds \mathcal O(x^{-d_0-2}) & \hbox{ near }x=\infty
 \end{array}
 \right .\\[20pt]
\ds \sqrt{\frac {\varkappa_j \tr(Y_j'(a_j)\sigma_+)}{\hbar(x-a_j)}}
(1+\mathcal O(x-a_j))\ \ \  \hbox{ near } x=a_j
 \end{array}\right.
\eea
This immediately implies the following identities
\bea
n\hbar +\frac 1 2  \sum_{r} t_{r,0} &\&= \mp \res{x=\infty}w_\pm dx\cr
\frac 1 2t_{0,J} &\&= \mp \res{x=\infty}\frac 1{x^J} w_{\pm} dx\ ,\ \
J\geq 1 \cr
\frac 1 2
t_{r,J} &\&=\mp \res{x=c_r} (x-c_r)^J w_\pm dx
\eea
and
\bea 
\hbar^2\frac{\pa}{\pa t_{r,J}} \ln \tau_n^{IM} &\&= \pm\res{x=c_r} (x-c_r)^{-j} w_\pm dx\label{ewq}
\\
\hbar^2\frac{\pa}{\pa c_{r}} \ln \tau_n^{IM}&\&= \pm\res{x=c_r} T'_r(x) w_\pm dx
\\
 \hbar^2\frac \pa{\pa a_j} \ln \tau_n^{IM}  &\&=\res{x=a_j} (w_\pm)^2dx\ .\label{qwe}
\eea
In order to compare with the formul\ae\ given in Thm. (\ref{eval}) we
note that the eigenvalues $w$ of $\mathcal A(x)$  and $Y$ of 
$\DD_n(x)$ are related as follows due to
 the change of gauge (\ref{newgauge})
\be
Y_\pm = \frac 1 2 V'(x) + w_\pm \ .
\ee 
Comparing eqs. (\ref{inf})--(\ref{aj}) with equations
(\ref{ewq})--(\ref{qwe}) we obtain
\beaq
\hbar^2 \pa_{t_{0,J}}\ln \frac {\ZZ_n}{\tau_n^{IM}}  &\&=  -\frac 1 {2J} \res{x=\infty}
x^J V'   \cr
 \hbar^2  \pa_{t_{r,J}}\ln \frac {\ZZ_n}{\tau_n^{IM}} &\&= - \frac 1 {2J} \res{x=c_r} 
\frac 1{(x-c_r)^j} V'(x)  dx  ,  \quad r=1,\dots,K,\ J=1,\dots,d_r   \cr
\hbar^2  \pa_{c_r}\ln \frac {\ZZ_n}{\tau_n^{IM}} &\&= -\frac 1 {2}\res{x=c_r} 
T'_r(x) V'(x)  dx  ,  \quad  r=1,\dots,K,   \cr
 \hbar^2  \pa_{a_j}\ln \ZZ_n &\&= \hbar^2  \pa_{a_j}\ln\tau_n^{IM}   \ .\label{tauzed}
\eeaq
These relations  define a closed differential 
\be
d\ln \left(\frac {\ZZ_n} {\tau_n^{IM}}\right) =:
    d \ln ( \FF_n)  \ ,\label{vacca}
\ee
where the  quantity $\FF_n$ is determined up to a multiplicative
factor independent of the  isomonodromic deformation parameters
$\{c_r, t_{rJ}^a, a_j\}_{J\ge 1}$, but which may depend on $n$.
This may be explicitly integrated to give  
\be
\ln \left( {\FF_n\over n!}\right)= {1\over 2 \hbar^2} \sum_{0\leq q<r\leq K} \res{x=c_r}
T'_r(x)T_q(x) \ ,\label{fudge}
\ee
where we have chosen to include the integration constant $\ln n!$ for reasons 
that will be explained in the remarks below.  Note that $\FF_n$ does not depend 
on the end-points parameters $\{a_j\}$, only those entering in the potential $V$.
Since the definition of the isomonodromic tau function $\tau^{IM}_n$
allows normalizations depending arbitrarily on the monodromy data we
obtain the following result.
\bt
\label{main}
Up to multiplicative terms that are independent of the isomonodoromic
deformation parameters $\{t_{r,J}, c_r, a_j\}$,
the partition function $\ZZ_n = \mathcal
Z_n(\{t_{r,J},c_r,a_j\}|[\varkappa])$ of the generalized random matrix
model and the isomonodromic tau function $\tau_n^{IM}$  for the
associated ODE are related by
\be
\ZZ_n =  \tau_n^{IM} \FF_n,
\ee
where $\FF_n$ is given in (\ref{fudge}).
\et

We conclude  with two remarks regarding the multiplicative factor 
$\FF_n$ relating $\ZZ_n$ and $\tau_n^{IM}$.

\br{(Change of  gauge)}

The factor $\FF_n$ can be  eliminated by another choice of gauge.
The original gauge is such that  near any singularity $c_r$ 
\be
\Psi(x) = C_r Y_r(x) {\rm e}^{-\frac 1 2 T_r (x)\sigma_3}\ .
\ee
(The point $x=\infty$  would require slight modifications in the argument which 
do not, however, change the result.)

The original tau function satisfies 
\be
\hbar^2d_{(r)} \ln \tau_n^{IM}  = -\frac 1 2 \res{c_r} \tr(Y_r^{-1}Y_r' \sigma_3 dT_r)\ .
\ee
We make a scalar  gauge transformation, depending on an arbitrary
parameter $c$, of the form
\be
\widetilde \Psi := \Psi {\rm e}^{- \frac c 2 V(x)} =C_r \widetilde Y_r 
   {\rm
  e}^{-\left(\frac 1 2\sigma_3 + \frac c 2 \1\right)T_r} 
\ee
where 
\be
\widetilde Y_r := Y_r(x){\rm e}^{-\frac c  2 f_r(x)}
\ee
and $f_r(x)$ is the regular part of $V(x)$ at the point $c_r$
\be
f_r(x) := \sum_{q\neq r} T_q(x)\ .
\ee
 The tau function for the gauge transformed system satisfies
\bea
d_{(r)} \ln \widetilde \tau_n^{IM}&\&  = \res{c_r}\tr \left(\widetilde Y_r^{-1}
\widetilde Y_r' \left(\sigma_3 - \frac c 2 \1\right) d T_r\right)\\
&\& =  \res{c_r}\tr \left[ \left(Y_r^{-1}
 Y_r'-\frac c 2 f_r'\right) \left(\sigma_3 - \frac c 2 \1\right) d T_r\right] \\
&\&  =
\frac {c^2}2 \res{c_r}f_r'dT_r +  \overbrace{\res{c_r}
\tr(Y_r^{-1}Y_r' \sigma_3 dT_r)}^{d_{(r)}\ln \tau_n^{IM}} - \frac c 2\res{c_r}\overbrace{\tr
  (Y_r^{-1}Y_r')}^{=f_r'}dT_r  \\
&\& =\frac {c^2}2 \res{c_r}f_r'dT_r + d_{(r)} \ln \tau_n^{IM} \ ,
\eea
where we have used that fact that
\be
\tr Y_r^{-1}Y_r' = 0\ .
\ee

Recall  that 
\be
d_{(r)} \ln \tau_n^{IM} = d_{(r)} \ln \ZZ_n - {1\over 2} d_{(r)} \ln \FF_n =  d_{(r)} \ln \ZZ_n  
+ {1\over 2} \res{c_r} f_r' dT_r
\ee
so that
\be
d_{(r)} \ln \widetilde \tau_n^{IM} =  \frac {c^2+1}2 
d_{(r)} \FF_n +   d_{(r)} \ln \ZZ_n \ .
\ee
It follows that if $c$ is chosen so that $c^2+1=0$ (i.e. $c = \pm i$) we have
$$
\widetilde \tau_n^{IM}  = \ZZ_n,
$$
up to a multiplicative factor independent of the 
isomonodromic deformation parameters.
\er
\br{(Schlesinger transformations)}
The shift $n\to n+1$ given by the ladder matrix $R_n(x)$ corresponds to an {\em elementary
  Schlesinger} transformation at infinity. (For more details on
Schlesinger transformation see \cite{JMII}.)

In general a Schlesinger transformation corresponds to a shift
by integers in  the spectrum of the logarithmic terms of the matrix
$T(x)$ entering the formal asymptotics.
 We see from formula (\ref{Gammacrasympt})
that the shift $n\to n+1$ amounts to increasing the first diagonal
entry of the logarithmic term  by $1$
and decreasing  the second entry by $1$. As shown in \cite{JMII} the two
corresponding tau functions are related by
\be
\frac{\tau^{IM}_{n+1}}{\tau^{IM}_{n}} = (Y_{0;1})_{12} \label{schl}
\ee
where the right hand side is the (1,2) matrix entry of the matrix
$Y_{0;1}$ in eq. (\ref{Trexp}). 
A simple computation (using
eq. (\ref{psinf})) shows that 
\be
 (Y_{n,0;1})_{12} = h_n,
\ee
in agreement with  Thm. \ref{main}, since (by
eq. (\ref{partit}))
\be
\frac {\ZZ_{n+1}}{\ZZ_n} = h_n (n+1)\ .
\ee
This explains why the integration constant $\ln n!$ has been included
in eq.~(\ref{fudge}); it assures that the dependence on the discrete
parameter $n$ in the relation between $\ZZ_n$ and $\tau_n^{IM}$ 
is consistent with that implied by the Schlesinger transformations.
\er
\bigskip


{\bf \large Acknowledgments}.
The authors would like to thank A. Borodin, M. Ismail and A. Kapaev
for helpful discussions.  This work was supported in part by the 
National Sciences Engineering Research Council of Canada.


\end{document}